\documentclass[twocolumn,times]{aastex631}

\usepackage{comment}

\shorttitle{SFE in gas-rich bars}
\shortauthors{Maeda et al.}
\graphicspath{{./}{figures/}}

\begin{document}

\title{Statistical Study of the Star Formation Efficiency in Bars: Is Star Formation Suppressed in Gas-Rich Bars?}

\author[0000-0002-8868-1255]{Fumiya Maeda}
\affiliation{Institute of Astronomy, Graduate School of Science, The University of Tokyo, 2-21-1 Osawa, Mitaka, Tokyo 181-0015, Japan}

\author[0000-0002-1639-1515]{Fumi Egusa}
\affiliation{Institute of Astronomy, Graduate School of Science, The University of Tokyo, 2-21-1 Osawa, Mitaka, Tokyo 181-0015, Japan}

\author[0000-0003-3844-1517]{Kouji Ohta}
\affiliation{Department of Astronomy, Kyoto University Sakyo-ku, Kyoto 606-8502, Japan}

\author[0000-0002-2107-1460]{Yusuke Fujimoto}
\affiliation{Department of Computer Science and Engineering, University of Aizu, Tsuruga Ikki-machi Aizu-Wakamatsu, Fukushima, 965-8580, Japan}

\author{Asao Habe}
\affiliation{Graduate School of Science, Hokkaido University, Kita 10 Nishi 8, Kita-ku, Sapporo, Hokkaido 060-0810, Japan}

\begin{abstract}
The dependence of star formation efficiency (SFE) on galactic {structures}, especially whether the SFE in the bar region is lower than those in the other regions, has recently been debated. We report the SFEs of 18 nearby gas-rich massive star-forming barred galaxies with a large apparent bar major axis ($\geqq 75^{\prime\prime}$). We statistically measure the SFE by distinguishing the center, bar-end, and bar regions for the first time. The molecular gas surface density is derived from archival CO(1--0) and/or CO(2--1) data by assuming a constant CO-to-H$_2$ conversion factor ($\alpha_{\rm CO}$), and the star formation rate surface density is derived from a linear combination of far-ultraviolet and mid-infrared intensities. The angular resolution is $15^{\prime\prime}$, which corresponds to $0.3 - 1.8~\rm kpc$. We find that the ratio of the SFE in the bar to that in the disk was systematically lower than unity (typically $0.6-0.8$), which means that the star formation in the bar is systematically suppressed. 
{Our results are inconsistent with similar recent statistical studies that reported that SFE tends to be independent of galactic structures.}
This inconsistency can be attributed to the differences in the definition of the bar region, spatial resolution, $\alpha_{\rm CO}$, and sample galaxies. Furthermore, we find a negative correlation between SFE and velocity width of the CO spectrum, which is consistent with the idea that the large dynamical effects, such as strong shocks, large shear, and fast cloud-cloud collisions caused by the noncircular motion of the bar, result in a low SFE.
\end{abstract}


\section{Introduction} \label{sec: intro}
Star formation activity changes within a galaxy and strongly depends on {galactic environments. Here, "environments" refer to structures within a galaxy such as spiral arms, bar, and nucleus in this paper.} In particular, a number of observations reported that massive star formation in the bar regions is suppressed in comparison with other regions. For instance, in the bar region of the strongly barred galaxies NGC~1300 and NGC~5383, the absence of prominent H\textsc{ii} regions (i.e., massive star formation) is reported \citep[e.g.,][]{tubbs_inhibition_1982,Sheth_NGC5383_2000}. In these galaxies, the molecular gas surface density ($\Sigma_{\rm mol}$) in the bar regions is  comparable to those in the bar-end and arm regions, where star formation is active \citep{maeda_large_2018}. This result indicates that the star formation efficiency (SFE $= \Sigma_{\rm SFR}/\Sigma_{\rm mol}$; where $\Sigma_{\rm SFR}$ is the star formation rate surface density) in the bar region is lower than those in other regions \citep{maeda_a_large_2020}. Furthermore, a low SFE in the bar region has been reported for other galaxies, including the bars of intermediate strength such as in NGC~2903 \citep{Muraoka_NGC2903_2016}, NGC~3627 \citep{law_submillimeter_2018}, 
NGC~4303 \citep{Momose_NGC4303_2010,yajima_co_2019}, 
NGC~4321 \citep{pan_variation_2017}, and
NGC~5236 \citep{Handa_bar_1991,Hirota_M83_2014}.

In contrast to the studies that observed the individual barred galaxies described above, recent statistical studies have suggested that the SFE in the bars is not systematically lower than those in other regions
and is, in fact, environmentally independent
\citep[e.g.,][]{Muraoka_radialSFE_2019,diaz-garcia_molecular_2021,Querejeta_stellar_2021}.
\citet{Muraoka_radialSFE_2019} measured the radial variations in the SFEs of  80 galaxies (30 SA, 33 SAB, and 17 SB galaxies) selected from the CO Multi-line Imaging of Nearby Galaxies (COMING) project \citep[][]{sorai_coming_2019}. The authors  found that the averaged SFEs of SA, SAB, and SB galaxies were nearly constant along the galactocentric radius.
\citet{Querejeta_stellar_2021} examined the environmental dependence in the SFEs of 74 galaxies (46 barred galaxies) from the Physics at High Angular resolution in Nearby GalaxieS ALMA (PHANGS-ALMA) project \citep[][]{leroy_phangsalma_2021}. 
The results of this study revealed that little difference existed in the SFEs among environments (center, bar, spiral arm, interarm, and disk without strong spiral). The authors did not find evidence of a systematically low SFE in the bar regions. Therefore, they concluded that galactic structures strongly affect the organization of molecular gas and star formation; however, their impact on SFE is small. \citet{diaz-garcia_molecular_2021} measured the SFE along the stellar bar of 12 strongly barred galaxies that host different degrees of star formation along the  major axis of the bar by using the IRAM 30~m single dish telescope.
They found that the SFEs are roughly constant along the bars and are not significantly different from the mean value in spiral galaxies that was reported by \cite{Bigiel_2011_KSlaw}. 

However, two methodological differences exist between the studies that focused on individual galaxies and the recent statistical studies. First difference is the definition of the bar. While the galactic center and bar-end are distinguished as other environments from the bar in most of the previous studies focused on individual galaxies, the definition of the bar in recent statistical studies includes (part of) center and bar-end, which may make the difference of SFEs between bars and disks small. This is because the SFE in the center and bar-end regions may be higher than that in the bar region as observed in some barred galaxies { \citep[e.g.,][]{Handa_bar_1991,Hirota_M83_2014,law_submillimeter_2018,yajima_co_2019,maeda_a_large_2020}.}

We emphasize that theoretical studies suggest clear distinction between center, bar, and bar-end in the star formation activity and gas dynamics.
Due to the non-axisymmetric gravitational potential,  some part of gas loses its angular momentum and falls to the center of the galaxy \citep[e.g.,][]{athanassoula_existence_1992}.
This gas inflow can create concentration of the molecular gas and induce active star formation in the center. The gas flows along the dust lane \citep[e.g.,][]{Regan_bar_1999}. It has been suggested that the non-circular gas motion in the bar region induces strong shock, large shear, and fast cloud-cloud collision, which can prevent molecular gas from forming stars \citep[e.g.,][]{tubbs_inhibition_1982,athanassoula_existence_1992,fujimoto_environmental_2014,fujimoto_fast_2020}. In the bar-end region,  continuous converging gas flow from the disk  and the bar regions causes gas accumulation, which can induce active star formation \citep[e.g.,][]{renaud_environmental_2015}.

{The second methodological difference between the studies that focused on individual galaxies and the recent statistical studies is spatial resolution.} The recent statistical studies included galaxies in the sample whose apparent bar major axis (i.e., the distance between both bar-ends) is as small as several times the angular resolution of the images they used, $15^{\prime\prime}-22^{\prime\prime}$. In this case, the center, bar, and bar-end cannot be distinguished, which may also smooth the differences in the SFE between the environments. 

In this paper, we aim to statistically determine whether the SFE in the bar region is lower than those in other regions  by distinguishing the galactic center, bar-end, and bar, similar to previous studies that focused on individual barred galaxies. To distinguish between these environments and avoid the smoothing the differences in the SFE, we focus on 18 gas-rich galaxies whose bar major axis is at least five times larger than the angular resolution of the images we used ($15^{\prime\prime}$). The sample selection and data reduction are presented in Section~\ref{sec: Sample and Data reduction}. 
As the main results of this paper, SFE profiles from the center to the bar and bar-end are presented in Section~\ref{sec: Results}.
In Section~\ref{sec: discussion}, we discuss  the effect of the angular resolution on the SFE profile, systematic uncertainties, and the relationship between the SFE and molecular gas properties (i.e., velocity width and line ratio). Finally, Section~\ref{sec: summary} presents a summary of this study.

\begin{figure}[t!]
\begin{center}
\includegraphics[width=\hsize]{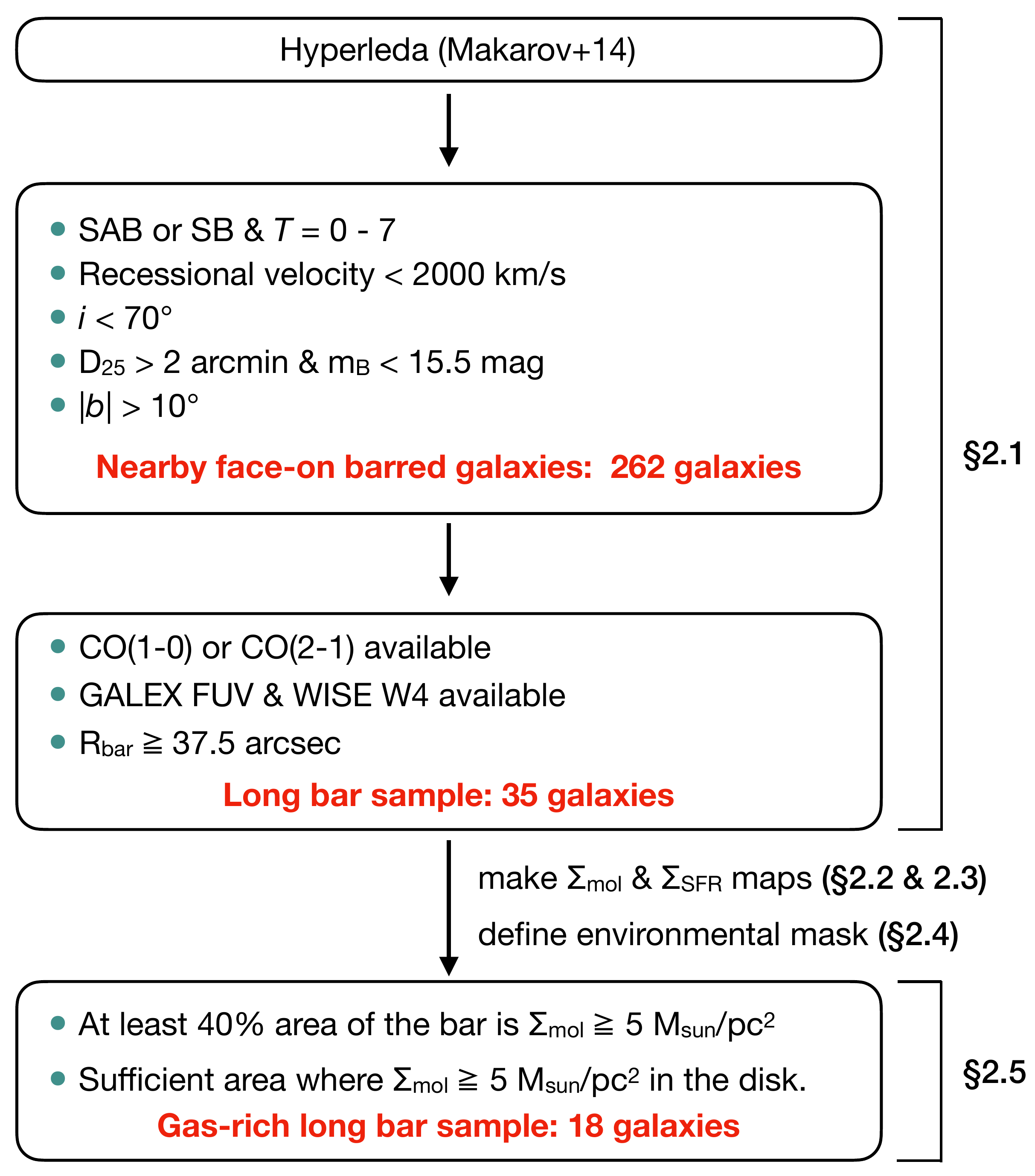}
\caption{Flowchart of our sample selection process.}
\label{fig: flowchart}
\end{center}
\end{figure}

\section{Sample and Data reduction} \label{sec: Sample and Data reduction}

This section describes the sample selection and images used in this study.
First, we select nearby face-on barred galaxies with available SFR and molecular gas tracers and with apparently long bar (Section~\ref{sec: long bar sample}). 
Then, we make $\Sigma_{\rm SFR}$ and $\Sigma_{\rm mol}$ maps (Sections~\ref{sec: Sigma_SFR} and \ref{sec: Sigma_mol}). In Section~\ref{sec: environmental mask}, we define the center, bar, bar-end, and disk regions. Next, in Section~\ref{sec: gas-rich long bar galaxies}, we select the galaxies with gas-rich bar and disk as the final sample. The final sample is referred to as the ``gas-rich long-bar sample'' in this study. Finally, in Section~\ref{sec: host gal.}, we describe the host galaxy properties of the gas-rich long-bar sample.
A flowchart of the sample selection process is presented in Figure~\ref{fig: flowchart}.

\subsection{Nearby face-on barred galaxies with an apparent long bar structure} \label{sec: long bar sample}

We first select 262 nearby face-on barred galaxies from the extragalactic database HyperLeda\footnote{\url{http://leda.univ-lyon1.fr}}\citep{Makarov_HyperLEDA_2014} according to the following criteria:

\begin{enumerate}
\item The morphological type is SAB or SB and the Hubble T stage is in $T = 0-7$, which corresponds to S0a--Sd. The morphological type is taken from the Third Reference Catalog of Bright Galaxies \citep[RC3;][]{de_Vaucouleurs_1991_rc3}.
\item The recessional velocity is $< 2000~\rm{km~s^{-1}}$ ($\sim~27~\rm Mpc$). The angular resolution of the images we use, 15$^{\prime\prime}$, which corresponds to a physical scale of less than 2.0 kpc.
\item The inclination ($i$) is  $< 70^\circ$. Because the inclinations taken from HyperLeda can be uncertain, when available, we adopt the inclinations from the catalogs of the PHANGS \citep{leroy_phangsalma_2021} and S$^4$G surveys \citep{sheth_spitzer_2010}. Here, we prioritize the PHANGS catalog over the S$^4$G catalog.
\item The projected major axis of a galaxy at the isophotal level 25 mag~arcsec$^{-2}$ in the $B$-band image ($D_{25}$) is $ > 2^\prime$, and  $B$-band magnitude is $< 15.5$~mag. 
This is used to remove  apparent small and/or faint galaxies.
\item The Galactic latitude ($b$) is $|b| > 10^\circ$. This is to minimize the contamination from the Milky Way disk.
\end{enumerate}

\begin{deluxetable*}{ccccccccccccccc}
\tablecaption{Properties of the long bar sample galaxies \label{tab:long bar galaxies}}
\tablewidth{0pt}
\tablehead{ 
\colhead{Galaxy}  & Morp.  & $D$ & $i$ & ref. & $\log M_\star$ & $\log {\rm sSFR}$  & $R_{\rm bar}$ &$\rm PA_{\rm bar}$ & $\epsilon_{\rm bar}$ & ref. & CO(1--0) & $T_{\rm rms}^{10}$ & CO(2--1) & $T_{\rm rms}^{21}$\\
 &    & (Mpc)  & ($^\circ$) &  & ($M_\odot$) & (${\rm yr}^{-1}$)   & ($^{\prime\prime}$)  & ($^\circ$)  & &  &  & (mK) & & (mK)
}
\decimalcolnumbers
\startdata
NGC~0613       & SBbc  & 25.1 & 35.7 & 1 & 10.96 & $-$10.03 &  85.1 & 125 & 0.75 & 5 & 7  & 46.2 & --& --  \\
{\bf NGC~1097} & SBb   & 13.6 & 48.6 & 2 & 10.76 & $-$10.08 &  95.1 & 141 & 0.65 & 5 & -- & --   & 2 & 0.3 \\
NGC~1291       & SB0a  &  8.6 & 29.3 & 1 & 10.60 & $-$11.33 &  95.1 & 165 & 0.41 & 5 & 10 & 23.2 & --& --  \\
NGC~1313       & SBd   &  4.3 & 34.8 & 2 &  9.55 &  $-$9.69 &  54.0 &  17 & 0.71 & 5 & -- & --   & 10& 2.1 \\
{\bf NGC~1300} & SBbc  & 19.0 & 31.8 & 2 & 10.50 & $-$10.44 &  84.4 &  99 & 0.78 & 5 & 10 &  6.8 & 2 & 0.8 \\
NGC~1317       & SABa  & 19.1 & 23.2 & 2 & 10.45 & $-$10.80 &  42.0 & 150 & 0.24 & 4 & 10 &  7.9 & 2 & 0.5 \\
NGC~1350       & SBab  & 20.9 & 64.6 & 1 & 10.73 & $-$11.05 &  55.7 &  36 & 0.58 & 5 & 10 &  6.3 & --& --  \\
{\bf NGC~1365} & SBb   & 19.6 & 55.4 & 2 & 11.06 &  $-$9.83 &  91.4 &  86 & 0.63 & 5 & 8  & 17.8 & 2 & 0.8 \\
NGC~1433       & SBab  & 18.6 & 28.6 & 2 & 10.68 & $-$10.65 &  86.2 &  95 & 0.65 & 5 & -- & --   & 2 & 0.9 \\
{\bf IC~0342}  & SABcd &  3.3 & 31.0 & 3 & 10.29 &  $-$9.67 & 120.0 & 155 & 0.42 &  3 & 3 & 34.8 & --& --  \\
NGC~1512       & SBa   & 18.8 & 42.5 & 2 & 10.60 & $-$10.49 &  73.5 &  42 & 0.66 & 5 & -- & --   & 2 & 0.6 \\
{\bf NGC~1672} & SBb   & 19.4 & 42.6 & 2 & 10.78 &  $-$9.90 &  69.3 &  96 & 0.67 & 5 & -- & --   & 2 & 0.8 \\
{\bf NGC~2903} & SABbc & 10.0 & 66.8 & 2 & 10.61 & $-$10.13 &  70.6 &  28 & 0.80 & 5 & 3  & 28.3 & 2 & 0.3 \\
NGC~3049       & SBab  & 30.8 & 58.0 & 1 &  9.97 &  $-$9.76 &  38.2 &  34 & 0.78 & 5 & -- & --   & 9 & 1.5 \\
NGC~3351       & SBb   & 10.0 & 45.1 & 2 & 10.28 & $-$10.18 &  53.6 & 112 & 0.46 & 5 & 3  & 28.6 & 2 & 0.8 \\
NGC~3359       & SBc   & 18.8 & 47.2 & 1 & 10.15 & $-$10.00 &  48.1 &  19 & 0.76 & 5 & 7  & 59.9 & --& --  \\
NGC~3368       & SABab & 10.9 & 51.1 & 1 & 10.55 & $-$10.75 &  62.8 & 124 & 0.43 & 5 & 7  & 48.0 & --& --  \\
{\bf NGC~3627} & SABb  & 11.3 & 57.3 & 2 & 10.78 & $-$10.20 &  59.1 & 160 & 0.76 & 5 & 3  & 32.7 & 2 & 0.4 \\
{\bf NGC~4051} & SABbc & 14.6 & 48.7 & 1 & 10.32 & $-$10.01 &  59.1 & 132 & 0.73 & 5 & 3  & 33.6 & --& --  \\
NGC~4258       & SABbc &  7.4 & 68.3 & 1 & 10.55 & $-$10.57 &  96.8 &   2 & 0.50 & 5 & 7  & 45.3 & --& --  \\
NGC~4293       & SB0a  & 15.8 & 65.0 & 2 & 10.36 & $-$10.64 &  70.6 &  75 & 0.75 & 5 & -- & --   & 2 & 0.8 \\
{\bf NGC~4303} & SABbc & 17.0 & 23.5 & 2 & 10.74 & $-$10.01 &  54.0 & 178 & 0.69 & 5 & 3  & 35.6 & 2 & 0.7 \\
{\bf NGC~4321} & SABbc & 15.2 & 38.5 & 2 & 10.72 & $-$10.17 &  54.6 & 108 & 0.59 & 5 & 3  & 10.8 & 2 & 0.5 \\
{\bf NGC~4535} & SABc  & 15.8 & 44.7 & 2 & 10.51 & $-$10.19 &  43.1 &  42 & 0.68 & 5 & 3  & 22.2 & 2 & 0.7 \\
NGC~4536       & SABbc & 16.3 & 66.0 & 2 & 10.44 &  $-$9.91 &  39.8 &  77 & 0.50 & 5 & 3  & 24.7 & 2 & 0.2 \\
{\bf NGC~4548} & SBb   & 16.2 & 38.3 & 2 & 10.52 & $-$10.80 &  58.0 &  61 & 0.53 & 5 & 3  & 24.5 & 2 & 0.5 \\
{\bf NGC~4579} & SABb  & 21.0 & 40.2 & 2 & 10.98 & $-$10.66 &  42.2 &  53 & 0.48 & 5 & 3  & 33.0 & 2 & 0.5 \\
NGC~4725       & SABab & 13.6 & 45.4 & 1 & 10.72 & $-$10.70 & 130.6 &  45 & 0.68 & 5 & -- & --   & 9 & 2.4 \\
NGC~4731       & SBcd  & 13.3 & 64.0 & 2 &  9.79 & $-$10.02 &  57.6 & 127 & 0.81 & 5 & -- & --   & 2 & 0.2 \\
NGC~4941       & SABab & 15.0 & 53.4 & 2 & 10.06 & $-$10.44 &  93.0 &  16 & 0.53 & 4 & -- & --   & 2 & 0.3 \\
{\bf NGC~5236} & SABc  &  4.9 & 24.0 & 2 & 10.52 &  $-$9.89 & 125.5 &  50 & 0.69 & 5 & 3  & 44.1 & 2 & 0.6\\
{\bf NGC~5457} & SABcd &  6.9 & 16.1 & 1 & 10.54 & $-$10.00 &  51.0 &  82 & 0.45 & 4 & 3  & 24.4 & 9 & 4.6\\
{\bf NGC~6946} & SABcd &  5.5 & 40.0 & 3 & 10.30 &  $-$9.79 &  60.0 &  17 & 0.46 & 4 & 3  & 26.3 & 9 & 3.3\\
{\bf NGC~6951} & SABbc & 24.1 & 30.0 & 3 & 10.73 & $-$10.19 &  44.0 &  88 & 0.54 & 6 & 3  & 13.5 & --& --  \\
{\bf NGC~7496} & SBb   & 18.7 & 35.9 & 2 & 10.10 &  $-$9.76 &  39.8 & 147 & 0.76 & 5 & -- & --   & 2 & 0.3
\enddata
\tablecomments{(1)~Galaxies with bold font indicate the gas-rich long-bar sample galaxies (Section~\ref{sec: gas-rich long bar galaxies}). (2)~Morphological type from RC3. (3)~Distance. (4)~Inclination. (5)~Reference of columns (3) and (4). (6)~Stellar mass. (7)~specific SFR. Derivation of columns (6) and (7) is described in Section~\ref{sec: long bar sample}. (8)~ Bar length. (9)~Position angle of the bar. (10)~Ellipticity of the bar. (11)~Reference of columns (8) - (10). (12)~Reference of CO(1--0) data. {(13)}~Median rms noise of the CO(1–-0) cube in $10~\rm km~s^{-1}$ bin. {(14)}~Reference of CO(2--1) data. {(15)}~Median rms noise of the CO(2–-1) cube in $10~\rm km~s^{-1}$ bin.\\
References: 1; S4G \citep{sheth_spitzer_2010}, 2; PHANGS-ALMA \citep{leroy_phangsalma_2021}, 3; Nobeyama CO atlas \citep{kuno_nobeyama_2007}, 4; \citet{menendezdelmestre_nearinfrared_2007}, 5; \citet{herrera-endoqui_catalogue_2015}, 6; \citet{diaz-garcia_molecular_2021}, 7; NRO COMING \citep{sorai_coming_2019}, 8; \citet{egusa_co_2022}, 9; HERACLES \citep{leroy_heracles_2009}, 10; from ALMA archival data (this work)}
\end{deluxetable*}

Among the 262 galaxies, we select those with available SFR and molecular gas tracers. Because we calculate the dust attenuation-corrected $\Sigma_{\rm SFR}$ by combining GALEX FUV and WISE 22~$\mu$m (W4; see Section \ref{sec: Sigma_SFR}), the galaxies without FUV or W4 image (35 galaxies out of 262) are excluded.
Then, we select the galaxies with CO(1--0) or/and CO(2--1) data cubes available. First, we refer to the following catalogs of previous CO survey projects: Nobeyama CO Atlas of Nearby Spiral Galaxies \citep[hereafter reffered to as NRO atlas;][]{kuno_nobeyama_2007}, COMING \citep{sorai_coming_2019}, HERA CO Line Extragalactic Survey \citep[HERACLES;][]{leroy_heracles_2009}, and PHANGS-ALMA \citep{leroy_phangsalma_2021}. For the galaxies outside these projects, we search the ALMA archival data sets by using the python package of \verb|astroquery|. We extract the galaxies with available mapping data from the ACA (7m+TP) since Cycle 1.  Here, the galaxies where CO mapping was only performed on a part of the disk were excluded (e.g., NGC~6744).
As a result, we extract 77 barred galaxies with available GALEX FUV, WISE W4, and CO(1--0) or/and CO(2--1) data sets.

Finally, from the 77 galaxies, we select those with an apparent long-bar structure. Bar structures are often defined as an ellipse which is defined by the center, semi-major axis (or bar-length; $R_{\rm bar}$), ellipticity ($\epsilon_{\rm bar}$), and position angle (${\rm PA}_{\rm bar}$). These parameters are calculated visually or based on an isophote with maximum ellipticity by using stellar images. In this study, we select the galaxies with $R_{\rm bar} \geq 37.^{\prime\prime}5$. This selection is based on the requirement that the major axis of the bar ($2R_{\rm bar}$) must be at least five times larger than the angular resolution of the tracers
 (i.e., $15^{\prime\prime}$; see Sections~\ref{sec: Sigma_SFR} and \ref{sec: Sigma_mol})  to distinguish between the center, bar, and bar-end.
For most galaxies, we adopt the catalog presented by \citet{herrera-endoqui_catalogue_2015}. The authors defined the bar parameters by using {\it Spitzer} 3.6$\mu$m images. 
For the galaxies outside the catalog, when available, we adopt the values reported in the literature that were calculated based on near-infrared images \citep{menendezdelmestre_nearinfrared_2007, kuno_nobeyama_2007, diaz-garcia_molecular_2021}.
For galaxies that did not have any corresponding literature, we visually ascertained whether $R_{\rm bar}$ is larger than $37^{\prime\prime}.5$; however, no galaxies are visually selected. 
Here, we manually exclude two galaxies (i.e., NGC~1055 and NGC~3556). These galaxies are listed in \citet{herrera-endoqui_catalogue_2015} as those with $R_{\rm bar} \geq 37^{\prime\prime}.5$. However, the true inclination appears to be nearly edge-on based on the visual inspection of the WISE and other optical images. 
{Consequently, we select 35 galaxies with $R_{\rm bar} \geq 37^{\prime\prime}.5$.}
These 35 galaxies are referred to as ``long bar sample galaxies'' in this paper. Table~\ref{tab:long bar galaxies} summarizes the properties of the long-bar sample galaxies.

\subsection{Star formation rate surface density} \label{sec: Sigma_SFR}
The $\Sigma_{\rm SFR}$ is calculated from a linear combination of GALEX FUV \citep{Gil_de_Paz_GALEX_2007} and WISE 22-$\mu$m \citep[W4;][]{Wright_WISE_2010} intensities, as reported by \citet{leroy_z_2019}.
\begin{eqnarray}
    \Sigma_{\rm SFR} &=& (\Sigma_{\rm SFR, FUV} +  \Sigma_{\rm SFR, W4}) \cos i \\
    \Sigma_{\rm SFR, FUV} &=& 1.04 \times 10^{-1} \left[ \frac{C_{\rm FUV}}{10^{-43.35}} \right] I_{\rm FUV} \\
    \Sigma_{\rm SFR, W4} &=& 3.24 \times 10^{-3} \left[ \frac{C_{\rm W4}}{10^{-42.7}} \right] I_{\rm W4},
\end{eqnarray}
where $\Sigma_{\rm SFR}$ is in units of $M_\odot~{\rm yr}^{-1}~{\rm kpc}^{-2}$, $I_{\rm FUV}$ and $I_{\rm W4}$ are the FUV and 22 $\mu$m intensities in units of $\rm MJy~sr^{-1}$, respectively, and conversion factors of $C_{\rm FUV}$ and   $C_{\rm W4}$ are  $10^{-43.42}$ and $10^{-42.73}~M_\odot~\rm yr^{-1}~(erg~s^{-1})^{-1}$, respectively. The systematic uncertainty translating from intensity to SFR is estimated to be $\approx 0.1$~dex.
Although the SFR can be calculated by using WISE 12~$\mu$m (W3) and GALEX NUV filters, we consider using FUV and W4 to be a better method for the following reasons. The coefficient to convert W3 intensity to the SFR estimated from spectral energy distribution fitting shows large cell-to-cell scatter \citep[e.g.,][]{leroy_z_2019} because the W3 filter  is prone to strong contaminations by the 11.3 $\mu$m polycyclic aromatic hydrocarbons (PAHs) features \citep{engelbracht_PAH_2005}. Additionally, the
NUV filter is prone to more contamination from lower mass and old stars compared with the FUV filter. The systematic uncertainties in $\Sigma_{\rm SFR, W4}$ are discussed in Section~\ref{sec: uncertainty of SFR}.

We use the delivered FUV and W4 images from the {$z=0$ Multiwavelength Galaxy Synthesis (z0MGS)} GALEX-WISE atlas data release 1 \citep{leroy_z_2019}, which are publicly available online\footnote{\url{https://irsa.ipac.caltech.edu/data/WISE/z0MGS/index.html}}. We use the image atlas, which consists of a set of background-subtracted images on matched astrometry with a matched resolution, at $15^{\prime\prime}$ resolution. In this study, the pixel size of the images are re-grided from the original size of $5.5^{\prime\prime}$ to half of the resolution of $7.5^{\prime\prime}$ by using the Python \verb|reproject| package.  {The $\Sigma_{\rm SFR}$ maps are shown in Figure~\ref{fig:FoV}(a). (The complete figure set (35 images) is shown in Appendix~\ref{ap: atlas}.)}

\subsection{Molecular gas surface density} \label{sec: Sigma_mol}

\subsubsection{CO(1--0) and  CO(2--1) data cubes}
Next, the $\Sigma_{\rm mol}$ is calculated from the CO(1--0) or/and CO(2--1) moment zero maps.
Columns (12) and (14) in Table~\ref{tab:long bar galaxies} summarize the references for the CO data cubes. The details of each reference are as follows.

{\it NRO atlas.} \citet{kuno_nobeyama_2007} presented CO(1--0) maps of 40 nearby spiral galaxies obtained from the Nobeyama 45 m telescope\footnote{\url{https://www.nro.nao.ac.jp/~nro45mrt/html/COatlas/}}. The beam size and noise levels are $15^{\prime\prime}$ and $40-100~\rm mK$ in a $5.0~\rm km~s^{-1}$ bin, respectively. For the galaxies observed in both the NRO atlas and COMING surveys, we prioritize the NRO atlas because of its higher angular resolution and sensitivity in comparison with the COMING.

{\it COMING.} \citet{sorai_coming_2019} presented  CO(1--0) maps of 147 nearby galaxies obtained from the Nobeyama 45 m telescope\footnote{\url{https://astro3.sci.hokudai.ac.jp/~radio/coming/}}. {The} beam size and noise levels are $17^{\prime\prime}$ and $\sim70~\rm mK$ in a $10.0~\rm km~s^{-1}$ bin, respectively. Although the sample number is approximately four times larger than that in NRO atlas, our target galaxies do not contain many COMING samples because COMING mainly targets the barred galaxies with $R_{\rm bar} < 37.^{\prime\prime}5$. According to \citet{yajima_R21_2021}, the gain uncertainty of the NRO atlas and COMING is estimated to be 25\%.

{\it HERACLES.} \citet{leroy_heracles_2009} presented  CO(2--1) maps of 48 nearby galaxies obtained from the IRAM 30 m telescope\footnote{\url{https://www2.mpia-hd.mpg.de/HERACLES/Overview.html}}. The beam size and noise levels are $13^{\prime\prime}$ and $20-25~\rm mK$ in a $2.6~\rm km~s^{-1}$ bin, respectively. The calibration uncertainty was reported to be $<20$~\% by \citet{den_brok_new_2021}. We use the cubes after convolving them to a beam size of $15^{\prime\prime}$.

{\it PHANGS-ALMA.} \citet{leroy_phangsalma_2021} presented  CO(2--1) maps of 90 nearby galaxies obtained from ALMA\footnote{\url{https://sites.google.com/view/phangs/home}}. The typical beam size and noise levels are $\sim1^{\prime\prime}$ and $\sim~\rm 85~mK$ in a $2.54~\rm km~s^{-1}$ bin, respectively. Most delivered data sets
include ACA (7 m array + total power) data. Therefore, the total flux was recovered. The gain uncertainty was estimated to be nominally 5~\%--10~\%. 
We use the delivered data cubes at a fixed angular resolution of $15^{\prime\prime}$.
For the galaxies observed by both the HERACLES and PHANGS-ALMA surveys, we prioritize PHANGS-ALMA because of its higher sensitivity in comparison with the HERACLES. 

{\it ALMA archival data.} CO(1--0) data of NGC~1291, NGC~1300, NGC~1317, NGC~1350, and NGC~1365, as well as CO(2--1) data of NGC~1313, are acquired from ALMA archival data.
For NGC~1365, CO(1--0) data were observed using the 12 m array and ACA under two projects namely, 2015.1.01135.S and 2017.1.00129.S. Using these data sets, \citet{egusa_co_2022} made a CO(1--0) cube with a beam size of $2.^{\prime\prime}0$ and median root mean square (rms) of 0.23~K. In this study, we use this cube after convolving it to a beam size of $15^{\prime\prime}$. Other galaxies were observed only by ACA under the projects of 2019.2.00052.S (NGC~1291), 2019.1.00722.S (NGC~1300), 2017.1.00129.S (NGC~1317 and NGC~1350), and 2018.A.00062.S (NGC~1313). For 7 m array data, we perform standard data reduction with CASA ver. 6.4.0 \citep{mcmullin_casa_ASPCS}. The total power data are added to the cleaned and primary-beam corrected 7 m array data via the CASA task \verb|feather|. Unfortunately, significant CO emissions are not observed in the data cubes of NGC~1291 and NGC~1313.

All data cubes were re-grided to the coordinate system of FUV and W4 images with a pixel size of $7.^{\prime\prime}5$ and were smoothed to a $10~\rm km~s^{-1}$ bin. The typical rms noise of each data cube is listed in Table~\ref{tab:long bar galaxies}.

\subsubsection{Conversion to $\Sigma_{\rm mol}$ }
First, we make velocity-integrated intensity maps (i.e., moment zero maps). For the spectrum of each line of sight (pixel), we identify consecutive channels, in which the signals are above $3\sigma_{\rm rms}$. For the data cubes not observed with ALMA, we adopt the median rms noise listed in Table~\ref{tab:long bar galaxies} as the $\sigma_{\rm rms}$. For the ALMA data, the $\sigma_{\rm rms}$ is calculated in each line of sight because {the} noise in the cube is nonuniform due to the primary beam pattern or/and difference in integration time within the field of view (FoV). We subsequently expand these channels to include all adjacent channels, in which the signals are above $1.5\sigma_{\rm rms}$. The velocity-integrated intensity of each pixel is defined as the sum of the masked channels; 
$\Sigma_{\rm mol}$ is derived from the CO(1--0) line as  follows:
\begin{equation}
    \Sigma_{\rm mol} = \alpha_{\rm CO} I_{\rm CO(1-0)} \cos i,
\end{equation}
where $\Sigma_{\rm mol}$ is in units of $M_\odot~\rm pc^{-2}$, $\alpha_{\rm CO}$ is the CO-to-H$_2$ conversion factor in units of $M_\odot~\rm (K~km~s^{-1}~pc^{2})^{-1}$, and $I_{\rm CO(1-0)}$ is the CO(1--0) velocity-integrated intensity in units of $\rm K~km~s^{-1}$. We adopt a constant Galactic $\alpha_{\rm CO}$ of  $4.35~M_\odot~\rm (K~km~s^{-1}~pc^{2})^{-1}$, including a factor of 1.36, to account for the presence of helium \citep{bolatto_conversion_2013}. 
We discuss the uncertainties in the $\alpha_{\rm CO}$ in Section~\ref{sec: alpha co}. 

From the CO(2--1) line, $\Sigma_{\rm mol}$ is derived by assuming the integrated intensity ratio ($R_{21}$) of CO(2--1)/CO(1--0) as follows:
\begin{equation}
    \Sigma_{\rm mol} = (\alpha_{\rm CO}/R_{21}) I_{\rm CO(2-1)} \cos i.
\end{equation}
Here, we adopt a constant $R_{21}$ of 0.65 based on recent  statistical studies on the ratio on a kpc scale in nearby galaxies \citep{den_brok_new_2021,Leroy_R21_2022}. We discuss the variation in the $R_{21}$ within the galaxy in Section~\ref{sec: R21}.
The $\Sigma_{\rm mol}$ maps from CO(1--0) and CO(2--1) are shown in Figure~\ref{fig:FoV}(c) and (d), respectively. (The complete figure set (35 images) is shown in Appendix~\ref{ap: atlas}.)

\begin{figure*}
\begin{center}
\figurenum{2}
\includegraphics[width=\hsize]{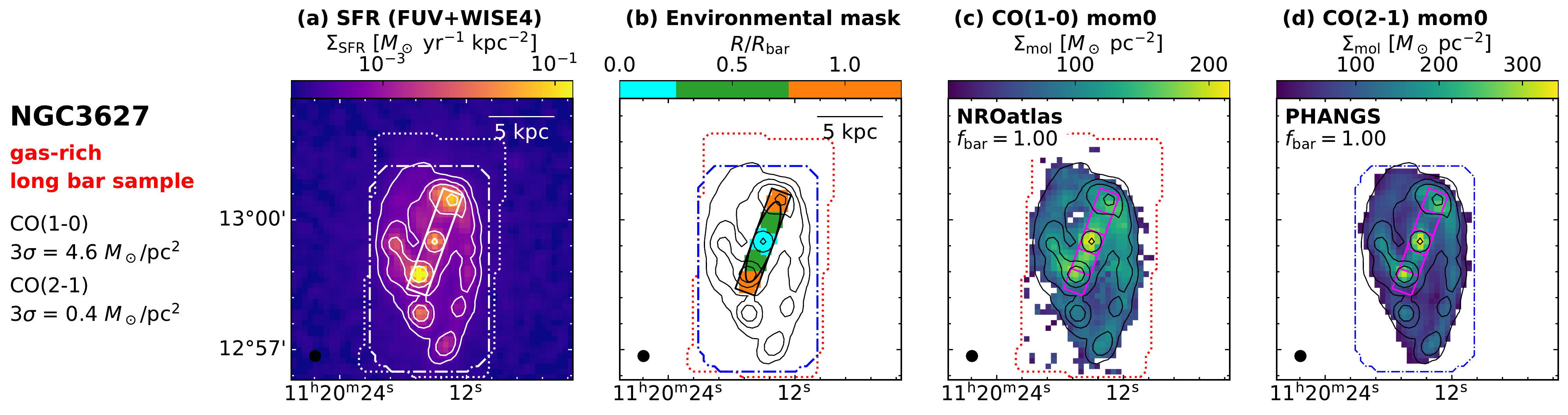}
\caption{NGC~3627. (a) $\Sigma_{\rm SFR}$ map derived from GALEX FUV and WISE 22$\mu$m. The contour levels are $\Sigma_{\rm SFR} = 10^{-2.5}, 10^{-2.0}, 10^{-1.5},$ and $10^{-1.0}~M_\odot~{\rm yr}^{-1}~{\rm kpc}^{-2}$. The FoVs of the CO(1--0) and CO(2--1) observations are represented as a white dotted and dash-dotted lines, respectively. The white rectangle is the environmental mask described in  Section~\ref{sec: environmental mask}. The black filled circle at the lower left corner represents the beam size of $15^{\prime\prime}\phi$.
(b) Environmental mask described in Section~\ref{sec: environmental mask}. The black ellipse is the cataloged bar structure (see Section~\ref{sec: long bar sample}). Color map shows the normalized distance to the minor axis of the ellipse. We defined the center, bar, and bar-end as the region where $R/R_{\rm bar} = 0.0 - 0.25$(cyan), $0.25-0.75$(green), and $0.75-1.25$(orange), in the rectangle, respectively.
(c) $\Sigma_{\rm mol}$ map derived from CO(1--0).
We display the region where $\Sigma_{\rm mol} \geqq 5~M_\odot~\rm pc^{-2}$.
The magenta rectangles represent the boundaries of the center, bar, and bar-end regions. {We show the $f_{\rm bar}(\Sigma_{\rm mol} \geqq 5)$ and surface density corresponding to $3\sigma$ upper limits of CO(1--0) cube. Here, the velocity width is assumed to be $20~{\rm km~s^{-1}}$.}
(d) $\Sigma_{\rm mol}$ map derived from CO(2--1). 
{(The complete figure set (35 images) is shown in Appendix~\ref{ap: atlas}.)}}
\label{fig:FoV}
\end{center}
\end{figure*}

\addtocounter{figure}{+1} 
\begin{figure*}[p]
\begin{center}
\includegraphics[width=165mm]{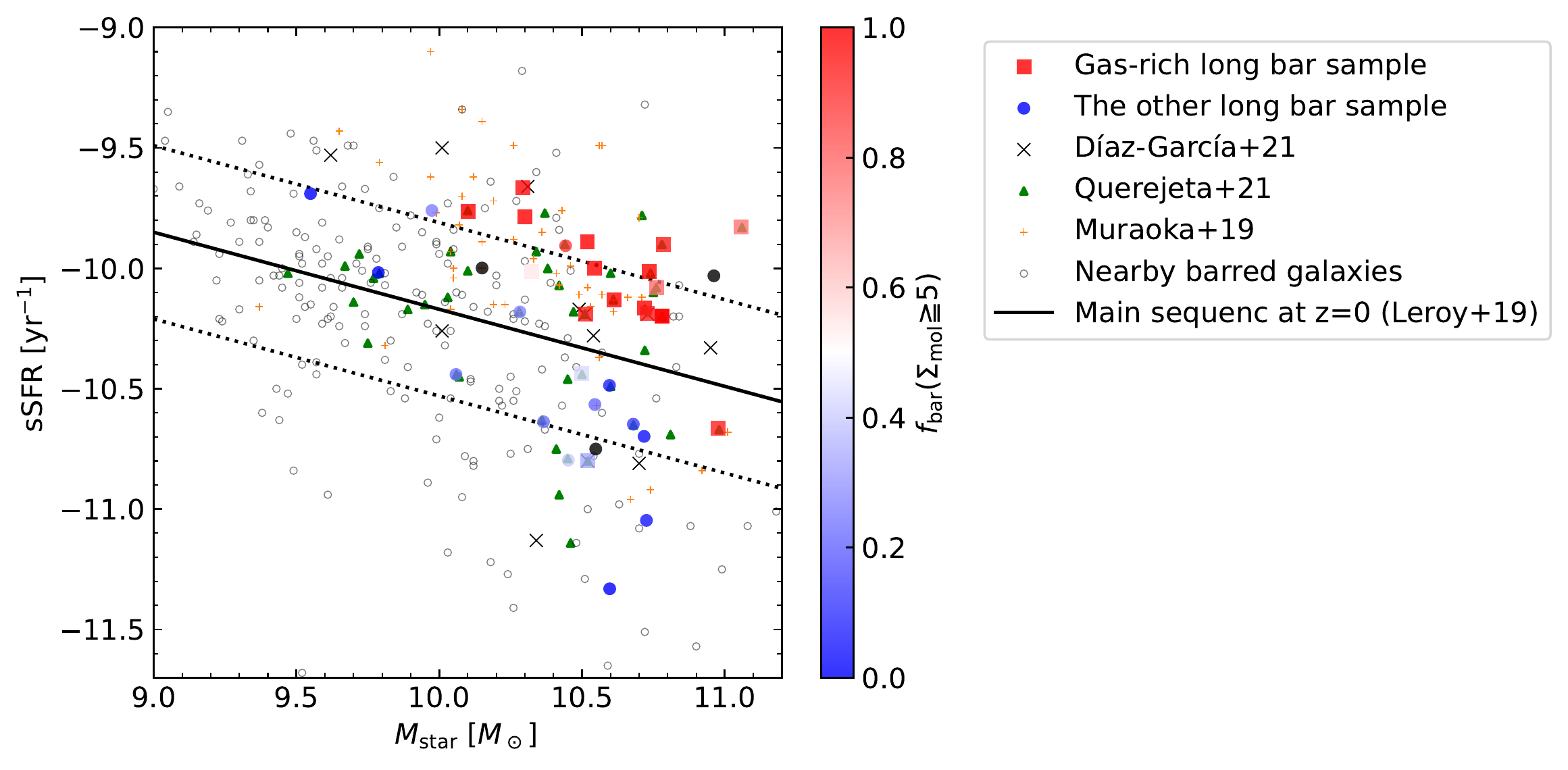}
\caption{
Distribution of the $f_{\rm bar}(\Sigma_{\rm mol} \geqq 5)$ of the long bar sample galaxies
on the $M_{\rm star}$ vs. sSFR diagram. 
Galaxies in the gas-rich long-bar sample are represented as square symbols. Other long bar sample galaxies are represented as circle symbols. 
The color of these symbols shows the $f_{\rm bar}(\Sigma_{\rm mol} \geqq 5)$.
Black filled circles show the galaxies with the high rms noise of 3$\sigma$ upper limit $\sim 10~M_\odot~\rm pc^{-2}$.
For galaxies with both CO(1--0) and CO(2--1) data cubes, the higher $f_{\rm bar}(\Sigma_{\rm mol} \geqq 5)$ is shown. 
The black crosses, green triangle, and orange pluses show the sample galaxies by \citet{diaz-garcia_molecular_2021}, \citet{Querejeta_stellar_2021}, and \citet{Muraoka_radialSFE_2019}, respectively. 
The black open circles show the nearby 262 barred galaxies described in Section~\ref{sec: long bar sample}.
Galaxies in the long bar sample galaxies and those of the above three studies are not shown for clarity. 
Black solid line shows the best-fitted main sequence by \citet{leroy_z_2019}. Scatter of their sSFR about this line is shown as black dotted lines. 
}
\label{fig:Mstar_vs_sSFR}
\end{center}
\end{figure*}

\subsection{Environmental mask} \label{sec: environmental mask}

The distinction between the center, bar, and bar-end is important because star formation activity is observably different  among these environments as described in Section~\ref{sec: intro}. 
In this study, we define the center, bar, and bar-end regions of the galaxy based on the stellar bar structure defined by the ellipse (see Section~\ref{sec: long bar sample}). We define a rectangle with a width of $2 \times 1.25 \times R_{\rm bar}$, height of $2R_{\rm bar} (1-\epsilon_{\rm bar})$, and position angle of $\rm PA_{bar}$, as shown in Figure~\ref{fig:FoV}.
In this rectangle, we define the center, bar, and bar-end as the regions where the distance to the minor axis of the ellipse ($R$) is $0.0 - 0.25R_{\rm bar}$, $0.25R_{\rm bar} - 0.75R_{\rm bar}$, and $0.75R_{\rm bar} - 1.25R_{\rm bar}$, respectively. The region outside of this rectangle and inside the FoV of the CO data cube is defined as a disk.
{The environmental mask maps are shown in Figure~\ref{fig:FoV}(b). (The complete figure set (35 images) is shown in Appendix~\ref{ap: atlas}.)}
Owing to the definition of the bar-end being $R/R_{\rm bar} = 0.75 - 1.25$, the peak of $\Sigma_{\rm SFR}$ around $R/R_{\rm bar} = 1.0$ and its surrounding region are excluded from the bar.
{The definition of the center being $R/R_{\rm bar} = 0 - 0.25$ also excludes the majority of bulge light from the bar; Referring to \citet{salo_spitzer_2015}, which measured the effective radius (half-light radius; $R_{\rm eff}$) of the bulge based on the S{\'e}rsic profile for S4G galaxies, we compare the $0.25R_{\rm bar}$ and $R_{\rm eff}$ for 23 galaxies in the long-bar sample. The $0.25R_{\rm bar}/R_{\rm eff}$ ranges from 1.4 to 10 with a median of 2.5. Therefore, the majority of bulge light of the long-bar sample seems to be included within the region defined as $0-0.25R_{\rm bar}$. }

As described in Section~\ref{sec: intro}, the definition of the bar region varied across studies. In many of the studies that focused on the individual barred galaxies, the bar region was defined as the region between the center and bar-end regions based on CO moment zero, optical, or near-infrared images \citep[e.g.,][]{Handa_bar_1991,Muraoka_NGC2903_2016,law_submillimeter_2018, yajima_co_2019,maeda_a_large_2020}. By contrast, some studies included  (part of) the center and bar-end in the region defined as bar \citep[e.g.,][]{Momose_NGC4303_2010}. In addition, star formation in the bar region was sometimes discussed using radial profiles \citep[e.g.,][]{James_Ha_2009,Hirota_M83_2014,Muraoka_radialSFE_2019}. Notably, our definition of the bar region does not include the center and bar-end regions, unlike recent statistical studies \citep{Querejeta_stellar_2021,diaz-garcia_molecular_2021}.

\begin{figure*}[t!]
\begin{center}
\includegraphics[width=\hsize]{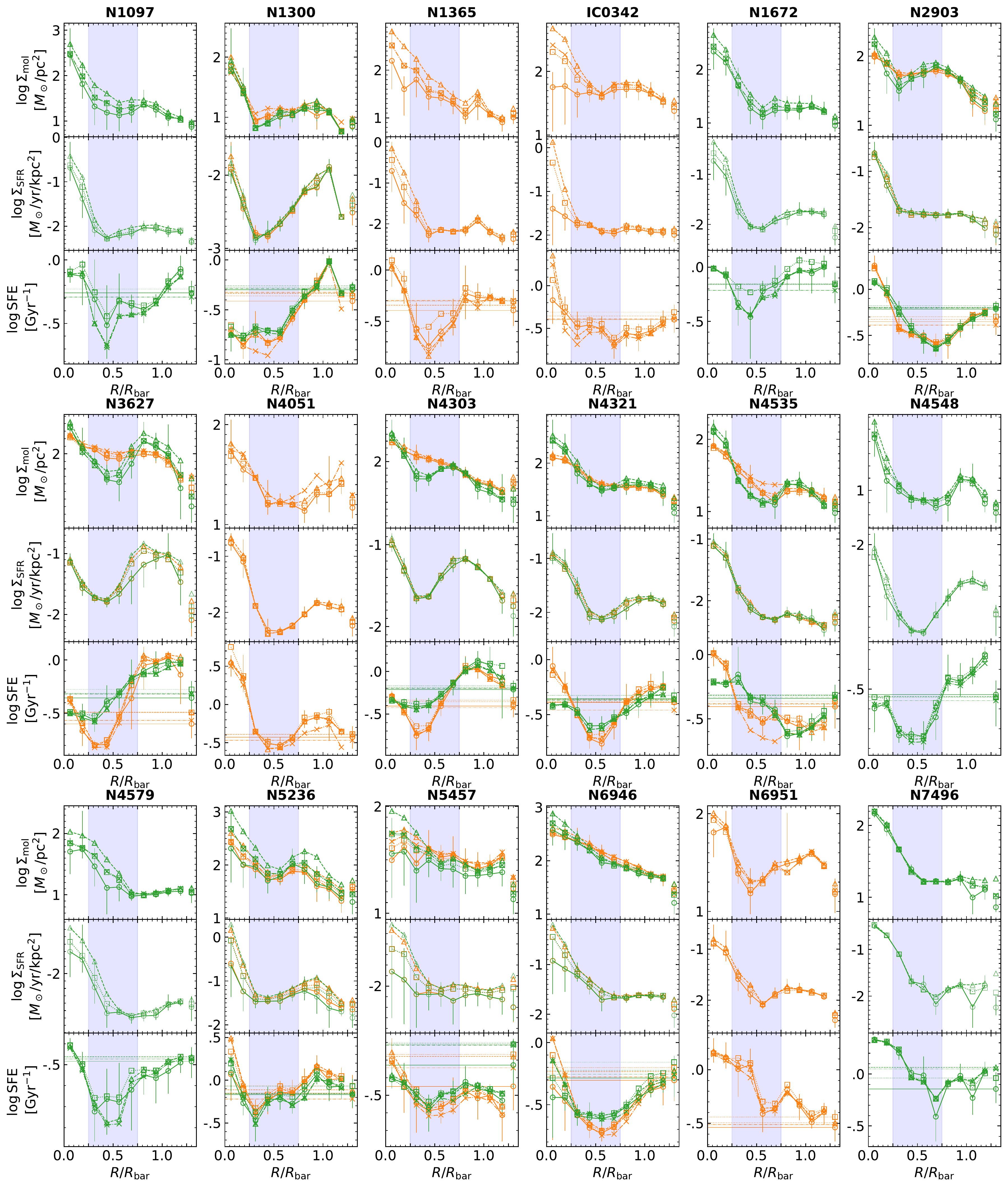}
\caption{Profiles of the $\Sigma_{\rm mol}$, $\Sigma_{\rm SFR}$, and SFE as a function of $R/R_{\rm bar}$ for the gas-rich long-bar sample galaxies. Orange and green represent the measurements by CO(1--0) and CO(2--1), respectively. The circle, square, triangle, and cross show median, mean, CO flux-weighted value, and value by the stacking method, respectively. 
The blue band shows the range of the bar region ($R/R_{\rm bar} = 0.25 - 0.75$).
The values in the disk are shown as symbols at $R/R_{\rm bar} = 1.3$. The SFEs in the disk are also shown as horizontal lines.
Note that the orange and green $\Sigma_{\rm SFR}$ profiles do not necessarily match because we do not used the same pixels in CO(1--0) and CO(2--1) data cubes.  }
\label{fig:Smol Ssfr SFE profile for each galaxy}
\end{center}
\end{figure*}

\subsection{Long bar galaxies with gas-rich bar and disk} \label{sec: gas-rich long bar galaxies}

To compare the SFE between the bar and disk regions, we select 18 galaxies with gas-rich bar and disk from the long-bar sample. 
In our CO data cubes, the rms noise exhibited a large variation. The rms noise is much higher in the CO(1--0) cube than in the CO(2--1) cube (see Table~\ref{tab:long bar galaxies} and Figure~\ref{fig:FoV}); typical surface densities corresponding to the $3\sigma$ upper limit of  CO(1--0) and CO(2--1) are $\sim5.0~M_\odot~\rm pc^{-2}$ and $\sim1.0~M_\odot~\rm pc^{-2}$, respectively. Therefore, for a fair comparison, we focus on the region where $\Sigma_{\rm mol} \geqq 5~M_\odot~\rm pc^{-2}$. According to this, we define a gas-rich bar as a bar region where
more than 40\% area is $\Sigma_{\rm mol} \geqq 5~M_\odot~\rm pc^{-2}$  in this study. This condition is represented as follows:
\begin{equation}
    f_{\rm bar}(\Sigma_{\rm mol} \geqq 5) \equiv A_{\rm bar}(\Sigma_{\rm mol} \geqq 5)/A_{\rm bar} \geqq 0.40,
\end{equation}
where $A_{\rm bar}(\Sigma_{\rm mol} \geqq 5)$ is the area of the region where $\Sigma_{\rm mol} \geqq 5~M_\odot~\rm pc^{-2}$ in the bar region, and  $A_{\rm bar}$ is the total area of the bar region. The $f_{\rm bar}(\Sigma_{\rm mol} \geqq 5)$ of each CO cube is shown in Figure~\ref{fig:FoV}. We select the CO(1--0) and CO(2--1) cubes that satisfy this condition. This condition allowed us to select the galaxies with a continuous distribution of molecular gas with $\Sigma_{\rm mol} \geqq 5~M_\odot~\rm pc^{-2}$ from the center to the bar-end region. We select 22 galaxies with a gas-rich bar from the long-bar sample galaxies. 

To compare the  SFEs between the bar and disk, we require sufficient pixels with $\Sigma_{\rm mol} \geqq 5~M_\odot~\rm pc^{-2}$ in the disk. By visual inspection, thus we exclude NGC~0613, NGC~1317, NGC~4536, and NGC~4941 as these galaxies have few or no pixels with $\Sigma_{\rm mol} \geqq 5~M_\odot~\rm pc^{-2}$ in the disk (see Figure~\ref{fig:FoV}). 
As for NGC~1365, we exclude only CO(2--1) due to the small FoV.
Consequently, we select 18 galaxies as gas-rich long-bar sample galaxies; these are represented in bold font in Table~\ref{tab:long bar galaxies}.
We use the CO(1--0) and CO(2--1) data cubes for 13 and 14 galaxies, respectively. We use both CO lines for the nine galaxies. 
The angular resolution of $15^{\prime\prime}$ corresponds to $0.3 - 1.8~\rm kpc$ of the gas-rich long-bar sample galaxies.
In the disks of most gas-rich long-bar sample galaxies, the regions where $\Sigma_{\rm mol} \geqq 5~M_\odot~\rm pc^{-2}$ corresponds to the regions where $\Sigma_{\rm SFR} \geqq 10^{-2.5}~M_\odot~\rm yr^{-1}~kpc^{-2}$. Although the 3$\sigma$ upper limit of the CO(1--0) cube is much higher than $5~M_\odot~\rm pc^{-2}$ in IC~0342, NGC~4303, and NGC~5236, they are included in the gas-rich long-bar sample because CO(1--0) is detected in most of the FoV. 
In the following section, we investigate the SFEs in the bar regions of the 18 gas-rich long bar sample galaxies.

\subsection{Host galaxy properties of gas-rich long-bar sample} \label{sec: host gal.}

Figure~\ref{fig:Mstar_vs_sSFR} shows the $f_{\rm bar}(\Sigma_{\rm mol} \geqq 5)$ of the target sample galaxies on the stellar mass ($M_{\rm star}$) vs. specific SFR (sSFR) diagram. The $M_{\rm star}$ and sSFR are listed in Table~\ref{tab:long bar galaxies}. The SFRs of the host galaxies are calculated by integrating  the region within the radius of $D_{25}$ in the $\Sigma_{\rm SFR}$ map. Similar to the SFR, $M_{\rm star}$ is calculated using the stellar mass surface density ($\Sigma_{\rm star}$) map. 
We follow the calculation of the $\Sigma_{\rm star}$ by \citet{leroy_z_2019} as $\Sigma_{\rm star} = 330 (\Upsilon_\ast^{3.4}/0.5) I_{\rm W1} \cos i$, where $\Sigma_{\rm star}$ is in units of $M_\odot~{\rm pc}^{-2}$, and $I_{\rm W1}$ is the intensities of the WISE 3.4~$\mu$m image  in units of $\rm MJy~sr^{-1}$. 
$\Upsilon_\ast^{3.4}$ is the near-infrared mass-to-light ratio in units of $M_\odot~L_\odot^{-1}$. We adopt $0.35~M_\odot~L_\odot^{-1}$, which is the average value in the PHANGS-ALMA sample \citep{leroy_phangsalma_2021}.

In Figure~\ref{fig:Mstar_vs_sSFR}, the galaxies in the gas-rich long-bar sample are represented by square symbols. The other long-bar sample galaxies are represented by circles.  We find that $f_{\rm bar}(\Sigma_{\rm mol} \geqq 5)$ depends on the $M_{\rm star}$ and sSFR; For barred galaxies with low stellar mass ($M_{\rm star} \leqq 10^{10}~M_\odot$) or those located in the lower side of the main sequence,  $f_{\rm bar}(\Sigma_{\rm mol} \geqq 5)$ is small ($< {0.4}$) and therefore, the $\Sigma_{\rm mol}$ in the bar tends to be low. By contrast, the galaxies with gas-rich bars ($f_{\rm bar}(\Sigma_{\rm mol} \geqq 5) > {0.4}$) tend to be located in the region where $M_{\rm star} \geqq 10^{10}~M_\odot$ and the upper side of the main sequence. Most of the gas-rich long-bar sample galaxies are located in this region. Such a dependence of the $f_{\rm bar}(\Sigma_{\rm mol} \geqq 5)$ is possibly related to the evolution of the barred galaxies, such as bar quenching process. Investigating this relationship can be interesting; however, this is beyond the scope of this study and can be a future study. The sample bias in our study is discussed in Section~\ref{sec: comparison with previous studies}.

\begin{figure*}[t!]
\begin{center}
\includegraphics[width=\hsize]{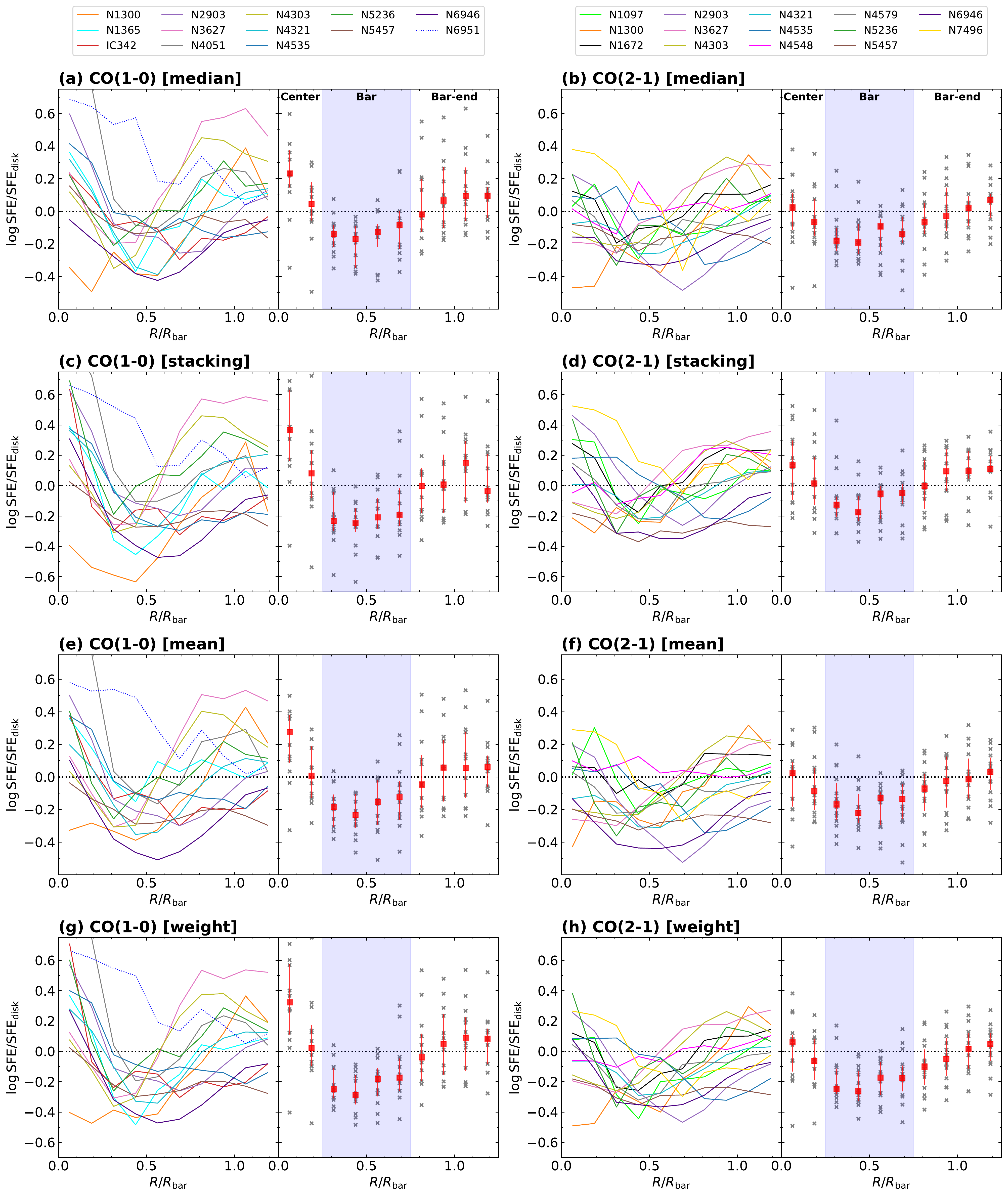}
\caption{ SFE profiles as a function of the $R/R_{\rm bar}$ of the gas-rich long-bar sample galaxies. (a) {Left side}  is the SFE profiles derived from CO(1--0). For each galaxy, we show the median SFE in each bin normalized by the median SFE in the disk. {Right side}  is the same as the {left plot}, but each data point is shown by gray cross and the median value and IQR of all gas-rich long-bar sample galaxies in each bin are shown as a red square and bar, respectively. Here, NGC~6951 is not included because of the large uncertainty (see Section~\ref{sec: gas-rich long bar galaxies}). The blue band shows the range of the bar region ($R/R_{\rm bar} = 0.25 - 0.75$).
(b) Same as panel (a), but for CO(2--1).
{(c)--(d) Same as panels (a) and (b), but SFE is derived by the stacking method (see text). 
(e)--(f) Same as panels (a) and (b), but median values are shown. 
(g)--(h) Same as panels (a) and (b), but CO flux-weighted values are shown. }
}
\label{fig: SFE profile medain stacking}

\end{center}
\end{figure*}

\begin{deluxetable*}{cc|cccc|cccc}
\tablecaption{Normalized SFE from $R/R_{\rm bar} = 0.00$ to $1.25$ \label{tab: result}}
\tablewidth{0pt}
\tablehead{ 
&& \multicolumn{4}{c|}{$\rm SFE/SFE_{disk}$ [CO(1--0)]} & \multicolumn{4}{c}{$\rm SFE/SFE_{disk}$ [CO(2--1)]} \\
Mask &$R/R_{\rm bar}$& median& stacking& mean& weighted &median& stacking& mean& weighted  
}
\decimalcolnumbers
\startdata
Center & $0.000-0.125$ & $1.70^{+0.86}_{-0.31}$ & $2.24^{+2.04}_{-0.78}$ & $2.00^{+0.51}_{-0.78}$ & $2.16^{+1.65}_{-0.87}$ & $1.02^{+0.39}_{-0.24}$ & $1.20^{+0.38}_{-0.44}$ & $1.07^{+0.30}_{-0.39}$ & $1.16^{+0.25}_{-0.41}$ \\
Center & $0.125-0.250$ & $1.11^{+0.54}_{-0.23}$ & $1.14^{+0.59}_{-0.31}$ & $1.02^{+0.67}_{-0.28}$ & $1.05^{+0.47}_{-0.25}$ & $0.90^{+0.35}_{-0.18}$ & $0.92^{+0.45}_{-0.25}$ & $0.88^{+0.34}_{-0.27}$ & $0.90^{+0.36}_{-0.25}$ \\
Bar & $0.250-0.375$ & $0.71^{+0.12}_{-0.12}$ & $0.59^{+0.32}_{-0.08}$ & $0.59^{+0.15}_{-0.10}$ & $0.56^{+0.23}_{-0.08}$ & $0.65^{+0.21}_{-0.07}$ & $0.61^{+0.14}_{-0.11}$ & $0.67^{+0.25}_{-0.13}$ & $0.59^{+0.14}_{-0.08}$ \\
Bar & $0.375-0.500$ & $0.68^{+0.13}_{-0.22}$ & $0.57^{+0.14}_{-0.07}$ & $0.56^{+0.23}_{-0.06}$ & $0.52^{+0.19}_{-0.06}$ & $0.63^{+0.20}_{-0.07}$ & $0.58^{+0.11}_{-0.10}$ & $0.63^{+0.11}_{-0.08}$ & $0.55^{+0.14}_{-0.09}$ \\
Bar & $0.500-0.625$ & $0.75^{+0.09}_{-0.18}$ & $0.62^{+0.21}_{-0.11}$ & $0.70^{+0.09}_{-0.19}$ & $0.66^{+0.12}_{-0.19}$ & $0.75^{+0.13}_{-0.16}$ & $0.66^{+0.19}_{-0.17}$ & $0.72^{+0.16}_{-0.19}$ & $0.63^{+0.19}_{-0.13}$ \\
Bar &$0.625-0.750$ & $0.83^{+0.17}_{-0.22}$ & $0.62^{+0.35}_{-0.07}$ & $0.71^{+0.18}_{-0.16}$ & $0.65^{+0.25}_{-0.09}$ & $0.75^{+0.07}_{-0.12}$ & $0.72^{+0.14}_{-0.20}$ & $0.74^{+0.18}_{-0.12}$ & $0.69^{+0.12}_{-0.16}$ \\
Bar-end &$0.750-0.875$ & $1.01^{+0.49}_{-0.27}$ & $0.95^{+0.37}_{-0.29}$ & $0.97^{+0.44}_{-0.32}$ & $0.93^{+0.38}_{-0.30}$ & $0.86^{+0.31}_{-0.12}$ & $0.81^{+0.32}_{-0.24}$ & $0.88^{+0.31}_{-0.26}$ & $0.79^{+0.31}_{-0.19}$ \\
Bar-end &$0.875-1.000$ & $1.22^{+0.66}_{-0.42}$ & $1.02^{+0.59}_{-0.35}$ & $1.20^{+0.48}_{-0.49}$ & $1.14^{+0.63}_{-0.44}$ & $0.99^{+0.37}_{-0.23}$ & $0.94^{+0.46}_{-0.28}$ & $0.97^{+0.46}_{-0.32}$ & $0.91^{+0.39}_{-0.22}$ \\
Bar-end &$1.000-1.125$ & $1.33^{+0.53}_{-0.46}$ & $1.41^{+0.55}_{-0.64}$ & $1.22^{+0.68}_{-0.48}$ & $1.27^{+0.42}_{-0.53}$ & $1.12^{+0.18}_{-0.32}$ & $1.14^{+0.34}_{-0.37}$ & $1.07^{+0.46}_{-0.29}$ & $1.10^{+0.27}_{-0.33}$ \\
Bar-end &$1.125-1.250$ & $1.21^{+0.22}_{-0.30}$ & $0.93^{+0.69}_{-0.16}$ & $1.06^{+0.25}_{-0.21}$ & $1.10^{+0.26}_{-0.27}$ & $1.20^{+0.26}_{-0.29}$ & $1.05^{+0.36}_{-0.17}$ & $1.14^{+0.34}_{-0.29}$ & $1.04^{+0.30}_{-0.19}$ \\
\enddata
\end{deluxetable*}

\begin{figure*}[t!]
\begin{center}
\includegraphics[width=170mm]{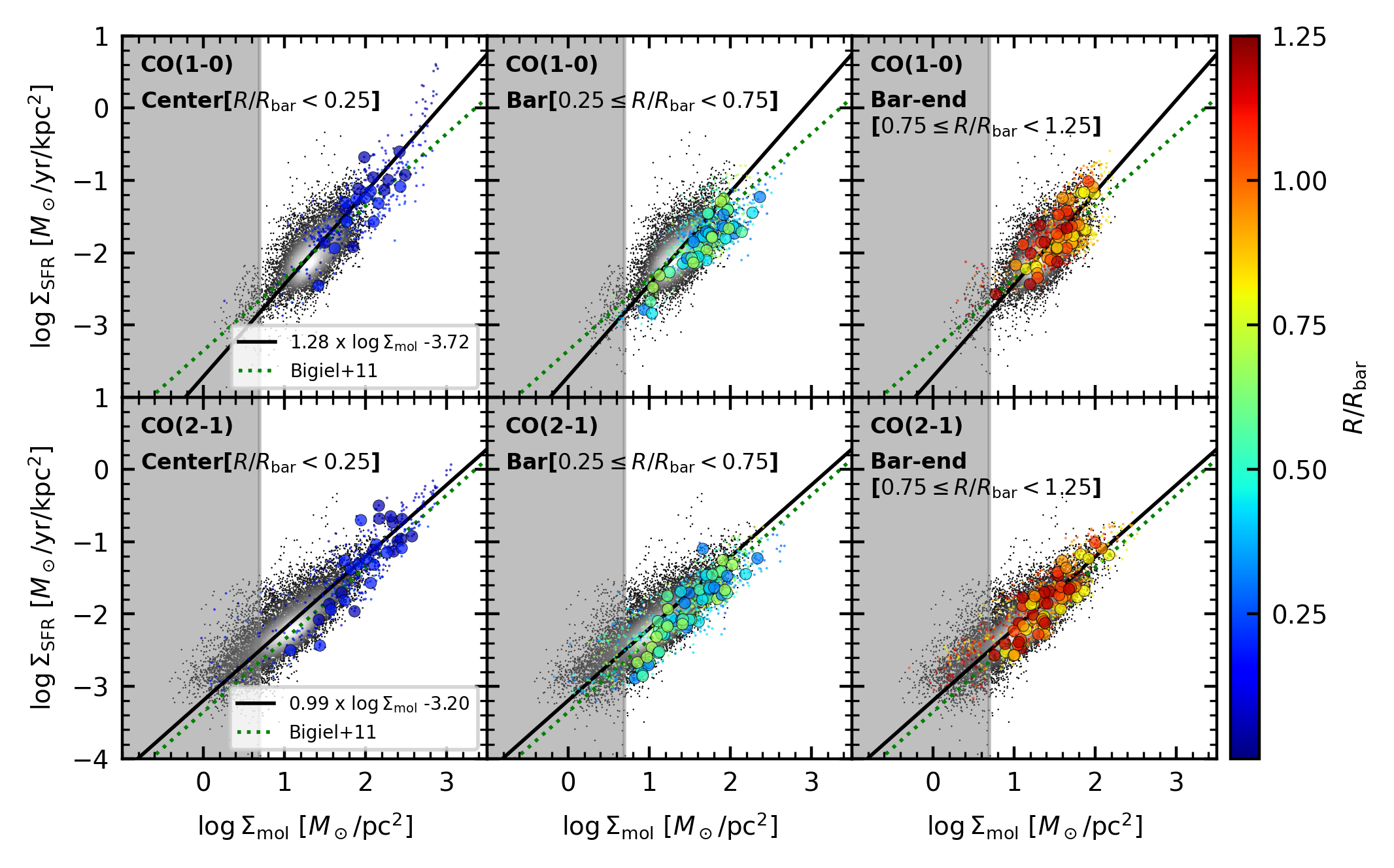}
\caption{Kennicutt-Schmidt relation of the gas-rich long-bar sample galaxies. Upper and lower panels show the relations using CO(1--0) and CO(2--1), respectively. Small colored points correspond to the pixel values in the center (left), bar (middle), bar-end (right). Large colored circles correspond to the median values in each $R/R_{\rm bar}$ bin. The gray data points correspond to the pixel values in the disk regions, whose best-fitted line is also shown with a black line. The green dotted line shows the best-fitted line obtained by \citet{Bigiel_2011_KSlaw}. The gray region correspond to $\Sigma_{\rm mol} \leqq 5.0~\rm M_\odot~\rm pc^{-2}$.}
\label{fig:KS law}
\end{center}
\end{figure*}

\section{Results} \label{sec: Results}

\subsection{Profiles of $\Sigma_{\rm mol}$, $\Sigma_{\rm SFR}$, and SFE} \label{sec: profile of Sigma_gas}

Figure~\ref{fig:Smol Ssfr SFE profile for each galaxy} shows the profiles of the $\Sigma_{\rm mol}$, $\Sigma_{\rm SFR}$, and SFE as a function of $R/R_{\rm bar}$ for the gas-rich long-bar sample galaxies. SFE is expressed as $\Sigma_{\rm SFR}/\Sigma_{\rm mol}$.  We divide $R/R_{\rm bar} = 0.0 - 1.25$ into 10 bins and derive the median (circle in the figure), mean (square), and CO flux weighted mean (triangle) in each bin. The bar with the circle shows the range of 75--25th percentile, which is known as the interquartile range (IQR).
The values in the disk region are shown at $R/R_{\rm bar} = 1.3$ for convenience. Additionally, the SFEs in the disk are represented using horizontal lines.
In addition, we obtained the $\Sigma_{\rm mol}$ by stacking the CO profiles in each bin, which are represented using  cross symbols. The stacking method followed in this study is the same as that in \citet[][refer to Section 3.3 for details]{maeda_a_large_2020}. In this case, the mean $\Sigma_{\rm SFR}$ is used in the SFE calculation.
In all gas-rich long-bar sample galaxies, significant variations are observed in the SFE from the center to the bar and bar-end; the SFEs tend to be higher at the center and bar-end than that at the bar. The dip in the SFE profile tends to be located at around $R/R_{\rm bar} = 0.5$, which  corresponds to the midpoint between the center and bar-end.
These results clearly demonstrate the importance of the distinction between the center, bar, and bar-end in SFE analysis. 

As reported in other studies  \citep[e.g.,][]{Leroy_R21_2022}, the NRO atlas data cubes suffer from visible mapping artifacts and poor baselines. Therefore, we compare the CO(1--0) fluxes from the NRO atlas with those from various independent observations, when available.
Seven galaxies namely, NGC~2903, NGC~3627, NGC~4303, NGC~4321, NGC~4535, NGC~5236, and NGC~6951 were compared; 
CO(1--0) was detected in the COMING project \citep{sorai_coming_2019} for NGC~2903, NGC~3627, and NGC~4303. The difference is within a factor of 1.5.
CO(1--0) in NGC~4321 was observed using ALMA under the project of 2011.0.00004.SV. in Cycle 0 as science verification data, and both the CO fluxes are comparable.
We confirm that the CO fluxes in the bar of NGC~5236 reported  by \citet{Hirota_M83_2014} and those in the NRO atlas are comparable.
We compare the CO fluxes in NGC~4535 and NGC~6951 with those reported by  \citet{diaz-garcia_molecular_2021}.
The differences in the CO fluxes in NGC~4535 is within a factor of 1.5. However, the CO fluxes in the bar of NGC~6951 reported by \citet{diaz-garcia_molecular_2021} are a factor of 6 higher than those in the NRO atlas. Owing to the lack of consensus on the CO(1--0) flux, NGC~6951 is not included in the remaining study. 
{We also compare the CO(2--1) fluxes between PHANGS-ALMA and HERACLES for the galaxies with both data available (i.e., NGC~2903, NGC~3627, NGC~4321, NGC~4579). The difference is within a factor of 1.2, which is consistent with the values reported by \citep{Leroy_R21_2022}.}

\subsection{Normalized SFE profile} \label{sec: normalixed SFE profie}

Figure~\ref{fig: SFE profile medain stacking} displays our main results, i.e., the SFE profiles from $R/R_{\rm bar} = 0.0$ to $1.25$ that are normalized by the SFE in the disk of the gas-rich long-bar sample galaxies.
The results obtained using CO(1--0) are shown in Figure~\ref{fig: SFE profile medain stacking}{(a), (c), (e), and (g)} and those obtained using CO(2--1) are shown in Figure~\ref{fig: SFE profile medain stacking}{(b), (d), (f), and (h).} Panels (a) and (b) show the median SFE in each bin that is normalized by the median SFE in the disk. The median values and IQRs of all the samples in each bin are shown as red squares and bars, respectively.
{The results when we use the SFE derived by the stacking method, the mean SFE, and  CO flux-weighted SFE as shown in panels (c) - (h).
Table~\ref{tab: result} summarizes the $\rm SFE/SFE_{disk}$  for each method.
This table shows the median value and IQR of all gas-rich long-bar sample galaxies in each $R/R_{\rm bar}$ bin, which correspond to red square and bar in Figure~\ref{fig: SFE profile medain stacking}, respectively.}

We find the SFE in the bar to be systematically lower than that in the disk regardless of whether  $\Sigma_{\rm mol}$ is measured using CO(1--0) or CO(2--1). The median normalized SFE ($\rm SFE/SFE_{disk}$) is $0.6 - 0.8$ in $R/R_{\rm bar} = 0.25-0.75$ regardless of the calculation methods. The
$\rm SFE/SFE_{disk}$ is at a minimum at around $R/R_{\rm bar} = 0.5$.
These results suggest that massive star formation in the bar region is systematically suppressed in comparison with the disk region in massive ($M_{\rm star} \geqq 10^{10}~M_\odot$) high sSFR (i.e., upper side of the main sequence) galaxies with gas-rich bar and disk. 
Our results are consistent with those reported in previous studies that observed individual barred galaxies \citep[][]{Handa_bar_1991,Momose_NGC4303_2010,Hirota_M83_2014,Muraoka_NGC2903_2016,pan_variation_2017,law_submillimeter_2018,yajima_co_2019,maeda_a_large_2020}.
Although the SFE in the bar region ($R/R_{\rm bar} = 0.25-0.75$) is systematically suppressed, its scatter is approximately 0.5~dex; some areas possess SFEs comparable to those in the disk, whereas others possess significantly lower SFEs than those in the disk. Additionally, the degree of suppression of star formation appears to vary among galaxies and within a galaxy.

In the center ($R/R_{\rm bar} = 0.0-0.25$), the SFE is higher or comparable to that in the disk. The $\rm SFE/SFE_{disk}$ that is obtained from CO(1--0) is higher than that obtained from CO(2--1). This is because $\Sigma_{\rm mol}$ in the center would be overestimated when we use the CO(2--1) line because of the high $R_{21}$ of $>0.65$ in the center (refer to Section~\ref{sec: R21}).
In the bar-end ($R/R_{\rm bar} = 0.75-1.25$), the $\rm SFE/SFE_{disk}$ are scattered in the range of  $1.0-1.3$ with peaks at $R/R_{\rm bar} \sim 1.0$. 
The star formation in the bar-end appeared to be slightly enhanced in comparison with that in the disk.

Although we focus on the region where $\Sigma_{\rm mol} \geqq 5~M_\odot~\rm pc^{-2}$, which corresponds to the typical 3$\sigma$ upper limit of the CO(1--0) data cubes, our conclusion does not depend on this threshold surface density. For the PHANGS galaxies in the gas-rich long-bar sample, we remeasure the $\rm SFE/SFE_{disk}$ by focusing on the region where $\Sigma_{\rm mol} \geqq 1~M_\odot~\rm pc^{-2}$, which corresponds to the typical 3$\sigma$ upper limit of the PHANGS data. As a result, the $\rm SFE/SFE_{disk}$ changes slightly. The differences are within 10~\%.

\subsection{Kennicutt-Schmidt relation}
The trend of obtaining a lower SFE in the bar region is additonally observed in the Kennicutt-Schmidt diagram or the $\Sigma_{\rm mol}$ vs. $\Sigma_{\rm SFR}$ diagram, as shown in Figure~\ref{fig:KS law}.
The large colored circles correspond to the median values in each $R/R_{\rm bar}$ bin for each galaxy.
The gray data points correspond to the pixel values in the disk regions, the best-fitted line of which is represented using a black line. We fit the data points in the disk region using the ordinary least-squares bisector method \citep{Isobe_fitting_1990}. The best-fitted relation is described as $\Sigma_{\rm SFR} = 10^{-3.72} \Sigma_{\rm mol}^{1.28}$ for CO(1--0) and $\Sigma_{\rm SFR} = 10^{-3.20} \Sigma_{\rm mol}^{0.99}$ for CO(2--1). These slopes are consistent with those reported in previous studies \citep[e.g.,][]{Bigiel_2011_KSlaw,yajima_R21_2021}. As shown in the middle panels, the median values in the bar regions are systematically located below the best-fitted lines, unlike those in the center and bar-end regions. Similar to Figure~\ref{fig: SFE profile medain stacking}, this result suggests that in the galaxies with gas-rich bar and disk, star formation tends to be suppressed in the bars in comparison with those in the disks.

\begin{figure*}[t!]
\begin{center}
\includegraphics[width=\hsize]{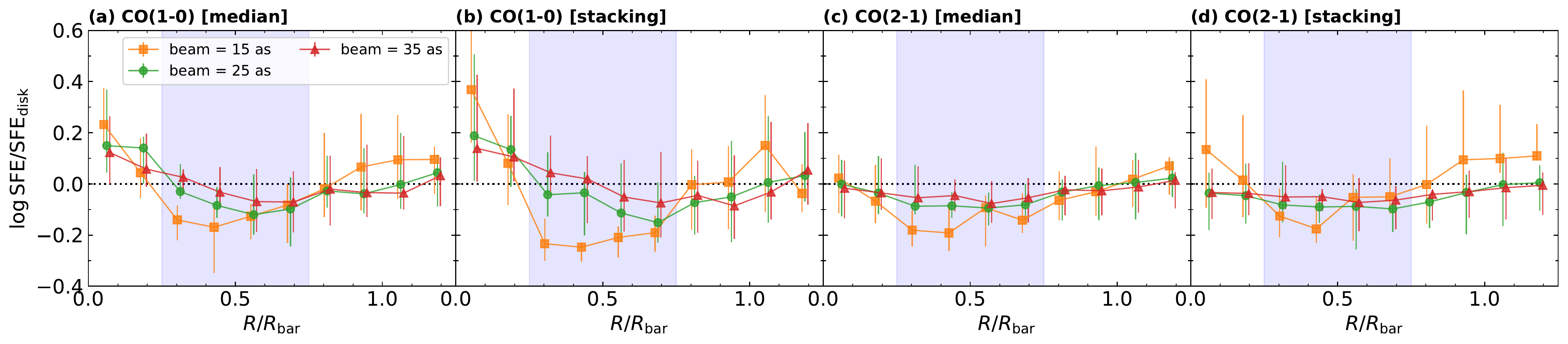}
\caption{Normalized SFE profiles when beam size is $15^{\prime\prime}$ (orange square), $25^{\prime\prime}$ (green circle) and $35^{\prime\prime}$ (red triangle). We show the median value and IQR in each $R/R_{\rm bar}$ bin.}
\label{fig: comparison beam size}
\end{center}
\end{figure*}

\begin{figure*}[tbhp!]
\begin{center}
\includegraphics[width=\hsize]{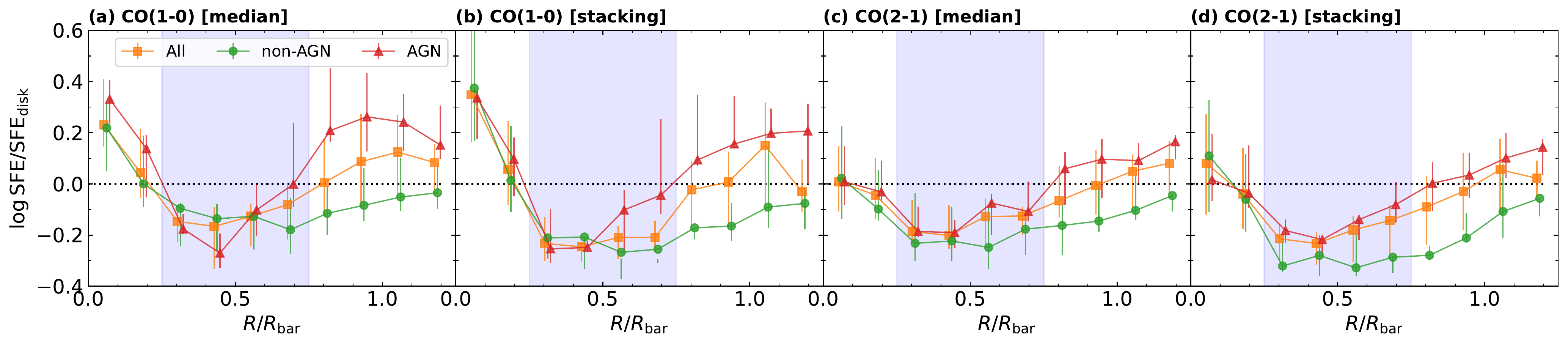}
\caption{Normalized SFE profiles of  all gas-rich long-bar sample (orange square),  non-AGN sample (green circle), and AGN sample (red triangle).
We show the median value and IQR in each $R/R_{\rm bar}$ bin.}
\label{fig: comparison AGV}
\end{center}
\end{figure*}

\section{Discussion} \label{sec: discussion}

\subsection{Beam size} \label{sec: beam size}
We investigate the effect of angular resolution on the SFE in the bar.
We remeasure the SFE using images convoluted to a beam size of $25^{\prime\prime}$ and $35^{\prime\prime}$.
In the case of beam size of $25^{\prime\prime}$ ($35^{\prime\prime}$), the bar lengths of approximately 50~\% (70~\%) galaxies in the gas-rich long-bar sample is less than five times the beam size.
Figure~\ref{fig: comparison beam size} displays the normalized SFE profiles when beam size is $15^{\prime\prime}$ (orange), $25^{\prime\prime}$ (green) and, $35^{\prime\prime}$ (red). 
Here, we use the pixels with $\Sigma_{\rm mol} \geqq 5~M_\odot~\rm pc^{-2}$ in the images with a beam size of $15^{\prime\prime}$.
As the beam size increases, the SFE profile is smoothed and become constant.
This result indicates that the SFE in the bars is possibly overestimated if the sample contains galaxies with bar length that is less than five times the beam size. 
One reason for the constant SFE radial profile or nonenvironmental dependence of SFE reported in the recent statistical studies 
would be that the beam size is large compared to the bar length of the sample galaxies.
In the studies by \citet{Muraoka_NGC2903_2016}, \citet{diaz-garcia_molecular_2021}, and \citet{Querejeta_stellar_2021},
the bar lengths of about 50, 80, and 40\% sample galaxies are less than five times the beam size of the images used in these studies ($17^{\prime\prime}$, $21.5^{\prime\prime}$, and $15^{\prime\prime}$, respectively). Therefore, the SFE radial profiles and maps would be smoothed and become constant.

\subsection{Systematic uncertainties} \label{sec: Uncertainties}

\subsubsection{CO-to-H$_2$ conversion factor} \label{sec: alpha co}
The choice of the $\alpha_{\rm CO}$ can be the largest source of uncertainty in measuring SFE.
The SFE ratio between the bar and disk depends on the ratio of the $\alpha_{\rm CO}$.
Therefore, if $\alpha_{\rm CO}$ varies within a galaxy, the profile of the $\rm SFE/SFE_{disk}$ may differ significantly from the profile obtained by assuming a constant $\alpha_{\rm CO}$.
Here, we discuss the potential for variations in the $\alpha_{\rm CO}$.

{\it Metallicity gradient.} 
As suggested by a number of studies \citep[e.g.,][]{Arimoto_xco_1996,Genzel_xco_2012,Accurso_xco_2017}, $\alpha_{\rm CO}$ increases with the decrease in  metallicity.  Considering the radial metallicity gradient within a galaxy \citep[e.g.,][]{Sanchez_metallicity_2014}, a radial gradient of $\alpha_{\rm CO}$ may be present.
Some studies in the PHANGS project \citep[e.g.,][]{sun_dynamical_2020,Querejeta_stellar_2021} used a metallicity-dependent conversion factor of  $\alpha_{\rm CO} \propto Z^{\prime -1.6}$ with an assumption of a radial metallicity gradient ($-0.1~{\rm dex}~R_{\rm e}^{-1}$; \citealp{Sanchez_metallicity_2014}), in which $Z^\prime$ is the metallicity that is normalized by the solar metallicity and $R_{\rm e}$ is the effective radius of the galaxy. 
This $\alpha_{\rm CO}$ gradient results in the $\rm SFE/SFE_{disk}$ in the bar being close to unity.
For the PHANGS sample galaxies in the gas-rich long-bar sample, we estimate the $\alpha_{\rm CO}$ using the same method as that used in the PHANGS project and find
that the $\alpha_{\rm CO}$ in the bar region is systematically $\sim0.15$~dex smaller than that in the disk.
Owing to this systematic difference, the difference in the SFEs between the bar and disk that is obtained by assuming a constant $\alpha_{\rm CO}$ almost disappears.
This result is consistent with the nonenvironmental dependence of  SFE reported by \citet{Querejeta_stellar_2021}.
(Notably, the authors also reported that the median SFE in the bar was approximately  0.7 times lower than that in the spiral arm when using the constant Milky Way $\alpha_{\rm CO}${; see also Section~\ref{sec: comparison with previous studies}). However, it may be more appropriate to assume a flat $\alpha_{\rm CO}$ rather than an $\alpha_{\rm CO}$ with a radial gradient because the radial gradient of the metallicity in barred galaxies is observed to be flatter than that in unbarred galaxies \citep[e.g.,][]{Martin_1994_radialZ,Zurita_2021_radialZ}, which would be caused by bar-induced mixing. The flat radial profiles of $\alpha_{\rm CO}$ reported in some barred galaxies \citep[e.g.,][]{Sandstrom_xco_2013,miyamoto_atomic_2021} support this picture.}

{\it Optically thin components.} 
Generally,  $\Sigma_{\rm mol}$ is calculated from the $^{12}$CO emission line using an $\alpha_{\rm CO}$ on the premise that the line is optically thick.
However, when the velocity gradient is large or/and the column density of the $^{12}$CO is small, $^{12}$CO emission line is optically thin.
In this case, using the standard $\alpha_{\rm CO}$ can overestimate the $\Sigma_{\rm mol}$.
In some bar regions, the presence of optically thin components was suggested.
In kpc resolution observations of 
$^{12}$CO(1--0) and $^{13}$CO(1--0) toward NGC~3627,
the integrated intensity line ratio of $^{12}$CO(1--0)/$^{13}$CO(1--0) in the bar region  was found to be as high as $\sim 20-30$ whereas the average value was $\sim 10$ \citep{watanabe_refined_2011, Morokuma-Matsui_NGC3627_2015}.
Here, $^{13}$CO(1--0) was assumed to be an optically thin line.
\citet{Morokuma-Matsui_NGC3627_2015} reported that such a high integrated intensity line ratio was possibly attributed to a high peak temperature ratio. The authors reported that both high integrated intensity ratio and high peak temperature ratio can not be explained under the assumption that $^{12}$CO(1--0) is optically thick, and suggested the presence of optically thin $^{12}$CO(1--0) components. 
Additionally, a higher $^{12}$CO(1--0)/$^{13}$CO(1--0) line ratio on a kpc scale in the bar region has been reported in other galaxies such as NGC~2903 \citep{Muraoka_NGC2903_2016} and NGC~4303 \citep{yajima_co_2019}.
Therefore, $\Sigma_{\rm mol}$  in the bar may be systematically overestimated in this study, and the SFE may be comparable to that in the disk.
However, the existence of such an optically thin component may be controversial because the above discussion is based on kpc resolution observations and under the assumption that $^{12}$CO and $^{13}$CO lines are emitted from the same cloud. The high-angular-resolution observations of both lines are important for further investigation.

\subsubsection{SFR} \label{sec: uncertainty of SFR}

The SFR derived from infrared (IR) emissions includes various contaminants.
One is the dust emissions from the old stellar population, which is known as IR cirrus.
The contribution from IR cirrus has been reported to be $30-60$~\% when $\Sigma_{\rm SFR}$ is less than $10^{-2}~M_\odot~\rm yr^{-1}~kpc^{-2}$ \citep{leroy_estimating_2012}.
On average the $\Sigma_{\rm SFR}$ that is less than  $10^{-2}~M_\odot~\rm yr^{-1}~kpc^{-2}$ obtained from GALEX FUV and WISE W4 are $20-30$\% higher than those obtained from Balmer-decrement-corrected H$\alpha$ in PHANGS-ALMA sample galaxies \citep{leroy_phangsalma_2021}.
Therefore, $\Sigma_{\rm SFR}$ in this study may be overestimated.
However, the IR cirrus does not seem to change our conclusion.
Because $\Sigma_{\rm SFR}$ is within similar ranges in the bar and disk regions (Figure~\ref{fig:KS law}), the contribution from IR cirrus is similar between the bar and disk regions, and the $\rm SFE/SFE_{disk}$ in the bar region is not expected to change significantly.

Another possible contamination is AGN, which would contributes to strong nuclear IR emission \citep[e.g.,][]{catalan-torrecilla_star_2015}.
The $\Sigma_{\rm SFR}$ in the center may be overestimated, although the contribution to $\Sigma_{\rm SFR}$ in the bar region is considered to be small because we select galaxies with an apparently large bar length.
We measure the SFE by distinguishing non-AGN from AGN in the gas-rich long-bar sample galaxies. 
The Seyfert or LINER that is extracted based on the catalogs by \citet{Ho_Seyfert_1997} and \citet{Veron_AGN_2010} are as follows: 
NGC~1097, NGC~1365, NGC~1672, NGC~3627, NGC~4051, NGC~4303, NGC~4321, NGC~4548, NGC~4579, NGC~6951, and NGC~7496.
Figure~\ref{fig: comparison AGV} shows the normalized SFE profiles of  non-AGN  and AGN samples.
The SFEs in the center of both samples are comparable, which suggests that the contamination by the AGN is small.
Regardless of the presence of AGN, the suppression of star formation is commonly observed. 
Interestingly, in the bar-end region, the $\rm SFE/SFE_{disk}$ tends to be $>0$ in the AGN sample and $< 0$ in the non-AGN sample.

\begin{figure*}[t!]
\begin{center}
\includegraphics[width=160mm]{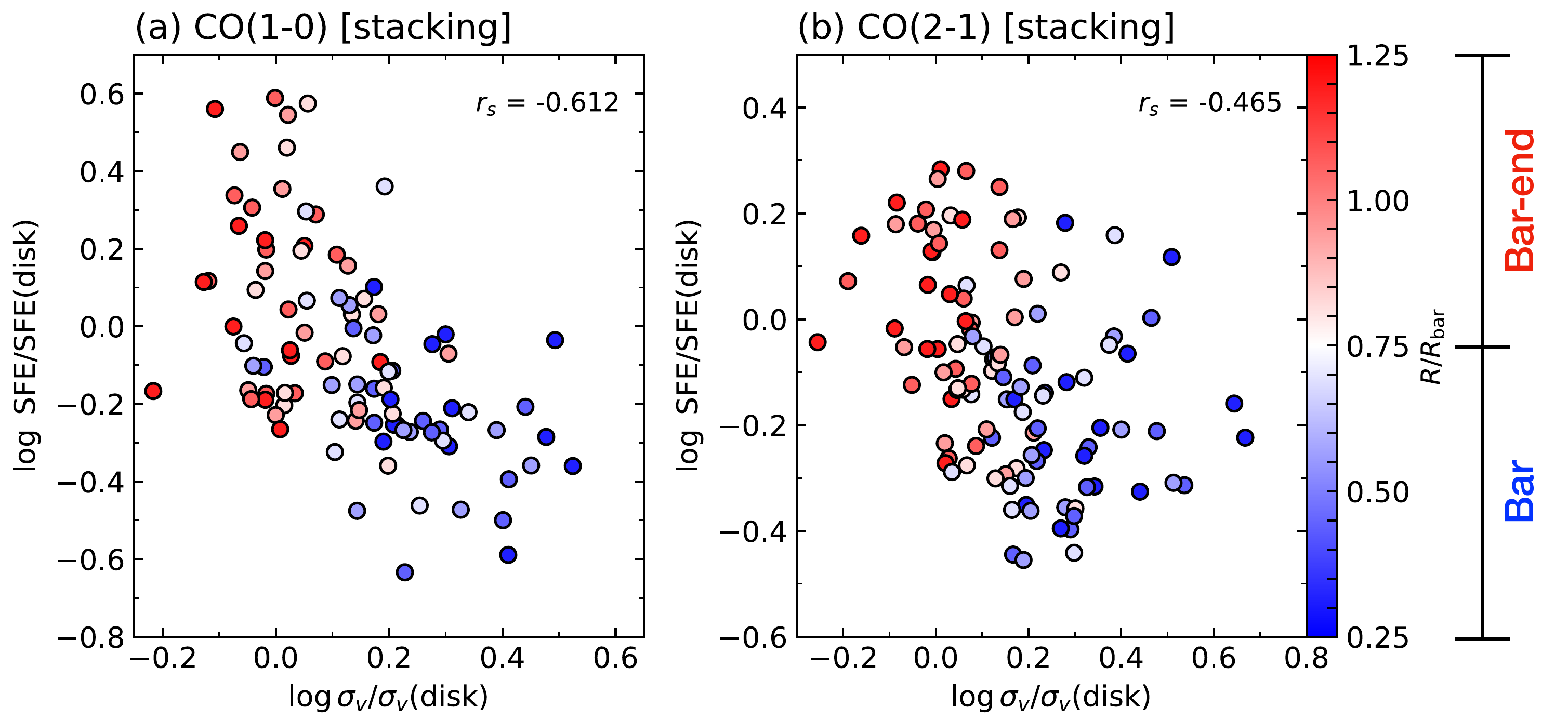}
\caption{(a) Relationship between normalized velocity width of the CO(1--0) spectrum and normalized SFE in the bar and bar-end regions of the gas-rich long-bar sample galaxies. The velocity width, which is derived from the stacking profile, is normalized by that in the disk region. The Spearman's correlation rank ($r_s$) is given in the top right corner. (b) Same as panel (a), but for CO(2--1). }
\label{fig:sigv vs SFE}
\end{center}
\end{figure*}

\subsection{Comparison with recent statistical studies} \label{sec: comparison with previous studies}

The environmental dependence of the SFE that is found in this study is inconsistent with that reported in recent statistical studies \citep{Muraoka_radialSFE_2019,Querejeta_stellar_2021,diaz-garcia_molecular_2021}.
Here, we summarize four possible causes for this inconsistency.

{\it (1) Bar definition.} 
In these studies, the (part of) center and bar-end regions are included in the region defined as the bar region, which would make SFE in their bars high. Our results that the $\rm SFE/SFE_{disk}$ in the center and bar-end regions is higher than that in the bar region (Figure~\ref{fig: SFE profile medain stacking}) support this possibility. In fact, the median $\rm SFE/SFE_{disk}$ in the range of $R/R_{\rm bar} = 0.0 -1.0$ in all the gas-rich long-bar sample is close to unity;
$0.91_{-0.09}^{+0.18}$ for CO(1--0) and $0.88_{-0.15}^{+0.05}$ for CO(2--1). 

{\it (2) Spatial resolution.} As described in Section~\ref{sec: beam size}, the bar length of about half of the sample galaxies in these studies is less than five times the beam size of the images they used, which would smooth the SFE profile.

{\it (3) CO-to-H$_2$ conversion factor.} 
{As already mentioned in Section~\ref{sec: alpha co}, \citet{Querejeta_stellar_2021} tested the impact of adopted $\alpha_{\rm CO}$ in SFE and found that the bar SFE becomes smaller with the constant $\alpha_{\rm CO}$ than that with the metallicity-dependent $\alpha_{\rm CO}$.
However, \citet{Muraoka_radialSFE_2019} and \citet{diaz-garcia_molecular_2021}  adopted the constant $\alpha_{\rm CO}$ and reported SFE being independent on environments. Therefore, the inconsistency with these two studies should be attributed to other factors than $\alpha_{\rm CO}$.}

{\it (4) Sample bias.} The difference between the sample galaxies in this study and the recent statistical studies possibly causes this inconsistency. 
We mainly focus on the galaxies located on the upper side of the main sequence with $M_{\rm star} \geqq 10^{10}~M_\odot$ in the $M_{\rm star}$ vs. sSFR diagram (Figure~\ref{fig:Mstar_vs_sSFR}).
However, the sample galaxies in the recent statistical studies are located within, as well as, outside the region as shown in the figure. 
Some sample galaxies in \citet{Muraoka_radialSFE_2019} posess a high sSFR ($\geqq 10^{-9.6}~\rm yr^{-1}$).
Many sample galaxies in \citet{Querejeta_stellar_2021} and \citet{diaz-garcia_molecular_2021} are located in the region where $M_{\rm star} \leqq 10^{10}~M_\odot$ and on the lower side of the main sequence. The SFE profiles of the galaxies located in these regions may differ from those of our gas-rich long-bar sample galaxies.
The SFE in the bar with a low stellar mass of $M_{\rm star} \leqq 10^{10}~M_\odot$ may be comparable to that in the disk.
The gas in the galaxies in the lower side of the main sequence may be depleted in the bar and disk, as shown by the blue symbols in Figure~\ref{fig:Mstar_vs_sSFR}, and star formation may be quenched in the entire disk, which may result in the SFE being constant. Investigating the relationship between the location of the host galaxies in the main sequence and changes in the SFE within galaxies will be important. CO and SFR tracers observations with high angular resolution and sensitivity are required for the further examination of the SFEs of low $M_{\rm star}$ and sSFR galaxies using the same methodology as that used in this study. This is because the apparent bar and disk sizes of low stellar mass galaxies are small and $\Sigma_{\rm mol}$ and  $\Sigma_{\rm SFR}$ of the galaxies located in the lower side of the main sequence are thought to be much lower than $5~M_\odot~\rm pc^{-2}$ and $10^{-2.5}~M_\odot~\rm yr^{-1}~kpc^{-2}$.

Note that the method of SFR calculation is not the source of the inconsistency because the SFR is derived from WISE W4 and GALEX FUV in all  statistical studies as in our study\footnote{\citet{diaz-garcia_molecular_2021} 
mainly used WISE W3 and GALEX NUV, but showed that their results does not change when using WISE W4 and GALEX FUV.}.

\begin{figure*}[thbp!]
\begin{center}
\includegraphics[width=\hsize]{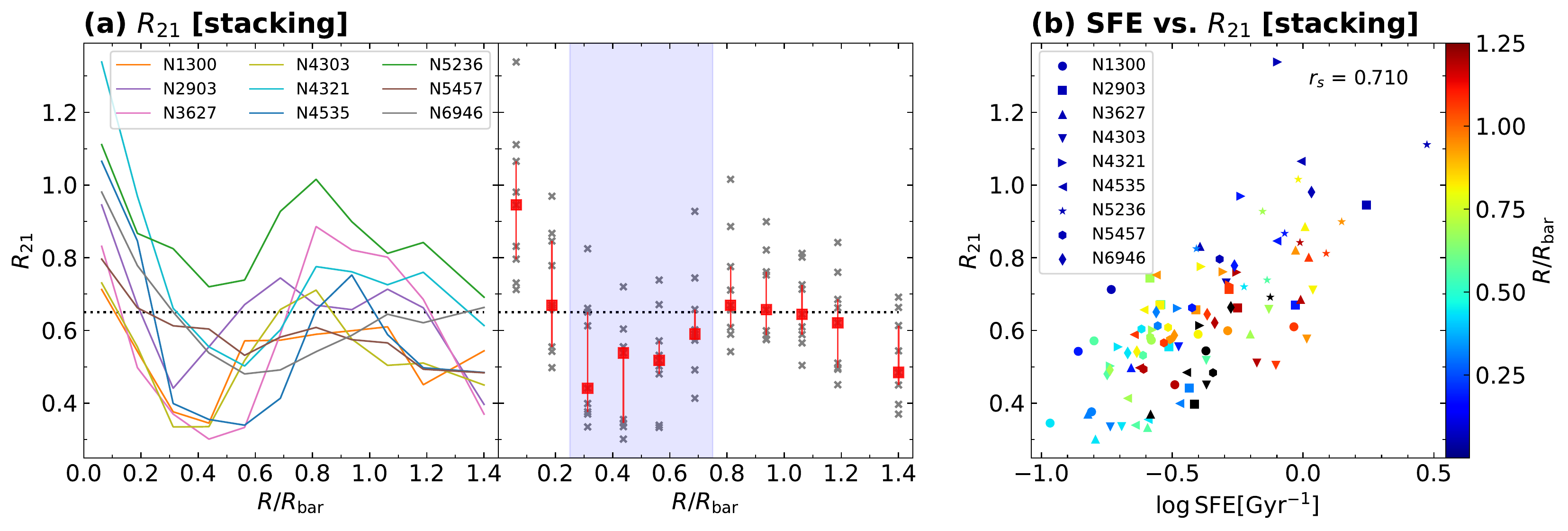}
\caption{(a) {Left side}  is the CO(2--1)/CO(1--0) line ratio profile. We use nine barred galaxies with available CO(1--0) and CO(2--1) data cube. 
The data points in the disk are shown at $R/R_{\rm bar} = 1.4$ for convenience.
{Right side} panel is the same as the {left plot}, but each data point is shown by gray cross and the median value and IQR of all sample in each bin are shown as a red square and bar, respectively. The black dotted horizontal line represents $R_{21} = 0.65$.
(b) Relationship between the SFE and the CO(2--1)/CO(1--0) line ratio. The Spearman's correlation rank ($r_s$) is given in the top right corner.
The data points in the disk are shown as black symbols for convenience. 
The typical error of the $R_{21}$ is estimated to be $\sim 25$~\%, which mainly contributed from the gain uncertainty for CO(1--0) data.
}
\label{fig: R21}
\end{center}
\end{figure*}

\subsection{Velocity width}

As mentioned in Section~\ref{sec: normalixed SFE profie}, the degree of the star formation suppression seems to vary among galaxies and within a galaxy. 
What determines the degree of the suppression? 
One promising parameter is the strength of noncircular motion in the bar region.
Star formation has been arguably suppressed by some dynamical effects, such as strong shock, large shear, and fast cloud-cloud collisions, which are  caused by the noncircular motion \citep[e.g.,][]{tubbs_inhibition_1982,athanassoula_existence_1992,Reynaud_Downes_bar_1998,emsellem_interplay_2015,fujimoto_environmental_2014,fujimoto_fast_2020,Maeda_CCC_2021}. 
Additonally, a number of CO observations have reported that the CO line width in the bar is larger than those in the bar-end and arm regions, which would support  the presence of large noncircular motion in the bar region \citep[e.g.,][]{Reynaud_Downes_bar_1998,Regan_bar_1999,watanabe_refined_2011,sorai_properties_2012,Morokuma-Matsui_NGC3627_2015,Muraoka_NGC2903_2016,maeda_large_2018,Sun_PHANGAS_2018,yajima_co_2019}.

Here, we investigate the relationship between the velocity width of the CO spectrum and SFE in the bar, bar-end, and disk regions of the gas-rich long-bar sample galaxies. Using the CO spectrum obtained by the stacking method described in Section~\ref{sec: profile of Sigma_gas}, we measure the effective width as a proxy for the line width.
We follow the definition of the effective width by \citet{Sun_PHANGAS_2018} as $\sigma_v = I_{\rm CO}/(\sqrt{2\pi} T_{\rm peak})$, where $T_{\rm peak}$ is the peak temperature of the spectrum. The effective width is less sensitive to noise in the line wings than in the second moment.

Figure~\ref{fig:sigv vs SFE} shows the relationship between the normalized velocity width of the CO(1--0) spectrum and normalized SFE in the bar and bar-end regions of the gas-rich long-bar sample galaxies. The velocity width is normalized by that in the disk region of the galaxy. We find negative correlations between the normalized velocity width and SFE, as indicated by the Spearman's rank correlation coefficient of $r_s = -0.612$ and $-0.467$ for CO(1--0) and CO(2--1), respectively. This trend is consistent with that reported by \citet{yajima_co_2019}, who investigated this relationship in NGC~4303.
The  $\sigma_v$ in the bar-end is $\sim0.8-1.2$ times larger than that in the disk, whereas that in the bar is  is $\sim1.2-4.0$ times larger than that in the disk. This negative correlation would support the idea that the larger the noncircular motion, the lower the SFE. 

{
This result can be interpreted as follows: In the bar-end regions, gas accumulates not only because of the stagnation of the gas in the elongated elliptical orbit due to the bar potential but also because of the inflow of the gas rotating in the disk \citep[e.g.,][]{downes_co_1996}, resulting in the moderate velocity width. Such orbital crowding increases the probability of cloud–cloud collision, leading to an increased gas density and the SFE \citep[e.g.,][]{renaud_environmental_2015}. In fact, the high gas density and SFE in the bar-end regions compared to those in the arm and bar regions  were reported in NGC~4303 \citep{yajima_co_2019}. Furthermore, frequent cloud-cloud collisions in the W43 GMC complex, which is considered to be located in the bar-end  of the Milky Way, are suggested \citep{Kohno_W43_2021}.}

{On the other hand, in the bar regions, the large velocity width suggests the presence of a strong shock, large shear, and fast cloud-cloud collisions compared to those in the bar-end region, leading to the low SFE. Some simulations showed that strong shock and/or large shear due to the non-circular gas motion by the bar potential destroy the molecular clouds and/or suppress the molecular cloud formation, leading to the suppression of the star formation \citep[e.g.,][]{tubbs_inhibition_1982, athanassoula_existence_1992, emsellem_interplay_2015, renaud_environmental_2015}. In fact, CO observations towards NGC~1530 suggested intense shock with high-velocity jumps and a large shear suppress the star formation by destroying the molecular clouds \citep{Reynaud_Downes_bar_1998}. In terms of the cloud-cloud collision, sub-pc scale simulations \citep[][]{takahira_cloud-cloud_2014,takahira_formation_2018} showed that a faster collision can shorten a gas accretion phase of the cloud cores formed, leading to suppression of core growth and massive star formation. Simulations of the cloud motion within a barred galaxy by \citet{fujimoto_fast_2020} showed that collision velocity between the clouds in the bar regions is larger than those in the other regions, which may be due to the perturbed motion of clouds to elliptical gas orbits  by gravitational interaction between clouds. Based on the sub-pc scale simulations, the authors proposed that fast collisions in the bar regions suppress the massive star formation \citep[see also][]{fujimoto_environmental_2014,fujimoto_giant_2014}.}

Because the velocity width is affected by the molecular gas distribution, relative velocities among molecular clouds, and gradient of the velocity field in the beam, {it is unclear  which of the above dynamical effects (shock, shear, and cloud-cloud collision) is dominant in the velocity width in the bar region based on the kpc-scale measurements only.}  Therefore, further observations at higher angular resolutions are required. For example, \citet{Maeda_CCC_2021} examined the motion of giant molecular clouds (GMCs) in NGC~1300 {on a spatial resolution of 40 pc} and found that the dispersion of the line-of-sight velocity among GMCs is larger in the bar than in the arm. Further, using the velocity field model, the authors suggested that the fast cloud-cloud collision in the bar region, which was caused by noncircular motion owing to the bar potential, suppressed star formation. Therefore, the large velocity width of the CO spectrum in the bar may reflect a fast cloud-cloud collision.

\subsection{CO(2--1)/CO(1--0) line ratio} \label{sec: R21}

The CO(2--1)/CO(1--0) line ratio, $R_{21}$  is dependent on the  gas conditions such as density and/or temperature, and the systematic variations in $R_{21}$ on a kpc scale have been observed in many galaxies \citep[e.g.,][]{leroy_heracles_2009,leroy_molecular_2013,Leroy_R21_2022,koda_physical_2012,koda_systematic_2020,Muraoka_NGC2903_2016,maeda_a_large_2020,den_brok_new_2021,yajima_R21_2021}.
Figure~\ref{fig: R21}(a) shows the $R_{21} (= I_{\rm CO(2-1)}/I_{\rm CO(1-0)})$ profiles of the gas-rich long-bar sample galaxies that have both CO(1--0) and CO(2--1) data available. 
Here, we stack the CO profiles of all pixels, in which both CO(1--0) and CO(2--1) lines are detected.
We observe environmental dependence; The $R_{21}$ in the center is the highest ($0.6-1.1$), followed by those in the bar-end ($0.6-0.8$) and bar ($0.4-0.6$).
Previous independent studies reported the same trend in the 
$R_{21}$ profiles of NGC~1300, NGC~2903, and NGC~5236 \citep{Muraoka_NGC2903_2016,maeda_a_large_2020,koda_systematic_2020}.
However, we emphasize that the range of $R_{21}$ in the disk is roughly comparable to that in the bar region, which suggests that the $\rm SFE/SFE_{disk}$ in the bar that was obtained from CO(2--1) does not strongly depend on $R_{21}$.

Figure~\ref{fig: R21}(b) shows the relationship between  SFE and  $R_{21}$. As indicated by the high $r_s$ of {0.710}, a strong positive correlation is observed. 
This result is consistent with that reported by \citet{maeda_a_large_2020},  who find the same trend in NGG~1300. The authors further find a negative correlation between the SFE and the fraction of diffuse extended molecular gas, which is missed in interferometer observations and would not directly contribute to the current star formation activity. They concluded that the SFE is roughly controlled by the amount of diffuse molecular gas. Our results would support this idea.
However, our results are inconsistent with that reported by \citet{Querejeta_stellar_2021}, who reported no clear trend in the $R_{21}$ from the center to the bar and bar-end and no correlation between the SFE and $R_{21}$.
The cause of these differences remains unclear. The possible causes includes  different sample galaxies and  spatial resolutions (Section~\ref{sec: comparison with previous studies}). Therefore, increasing the sample number of $R_{21}$ maps with a sufficient angular resolution to resolve the environments will be important.

\section{Summary} \label{sec: summary}
We statistically investigate the SFE {variation} within the galaxy by focusing on 18 nearby face-on gas-rich barred galaxies with large apparent bar lengths ($R_{\rm bar} \geqq 37^{\prime\prime}.5$). Most of the 18 galaxies are massive ($M_{\rm star} \geqq 10^{10}~M_\odot$) and located in the upper side of the main sequence  (Figure~\ref{fig:Mstar_vs_sSFR}). Unlike similar recent statistical studies, we measure the SFE by distinguishing between the center, bar-end, and bar (i.e., the ridge region between the center and the bar-end) for the first time. The $\Sigma_{\rm SFR}$ is derived from the linear combination of GALEX FUV and WISE 22$\mu$m intensities, and the $\Sigma_{\rm mol}$ is derived from the CO(1--0) or/and CO(2--1) lines by assuming a constant $\alpha_{\rm CO}$.
The angular resolution is $15^{\prime\prime}$, which corresponds to $0.3 - 1.8~\rm kpc$.
We focus on the region where $\Sigma_{\rm mol} \geqq 5~M_\odot~\rm pc^{-2}$.
The main results obtained are as follows:

\begin{enumerate}
   \item In all 18 galaxies, significant variations in the SFEs from the center to the bar and bar-end are observed. The SFEs tend to be higher in the center and bar-end, and lower in the bar. The dip in the SFE profile tends to be located at around $R/R_{\rm bar} = 0.5$ (Figure~\ref{fig:Smol Ssfr SFE profile for each galaxy}). 
   
   \item The SFE in the bar region is found to typically be $0.6-0.8$ times lower than that in the disk region, which suggests that the star formation is systematically suppressed. The SFEs in the center and  bar-end are higher or comparable to that in the disk (Figure~\ref{fig: SFE profile medain stacking}).
   
   \item Although the SFE in the bar region is systematically suppressed, the ratio of the SFE in the bar region to that in the disk exhibits a scatter of approximately 0.5~dex. The degree of star formation suppression varies among galaxies and within a galaxy.
   
   \item Our results are inconsistent with the results of nonenvironmental dependence on the SFE that is obtained by similar recent statistical studies \citep[][]{Muraoka_radialSFE_2019,Querejeta_stellar_2021,diaz-garcia_molecular_2021}. The possible causes of this inconsistency are the differences in the definition of the bar region, spatial resolution, the $\alpha_{\rm CO}$, and sample galaxies (Sections~\ref{sec: beam size}, \ref{sec: Uncertainties}, and \ref{sec: comparison with previous studies}).
   
   \item We find a negative correlation between the SFE and velocity width of the CO spectrum, which would support the idea that the strength of non-circular motion controls the degree of the star formation suppression (Figure~\ref{fig:sigv vs SFE}).
   
   \item We find a positive correlation between the SFE and the CO(2--1)/CO(1--0) ratio, which would support the idea that SFE is roughly controlled by the amount of diffuse molecular gas (Figure~\ref{fig: R21}).

\end{enumerate}

In conclusion, our results clearly demonstrate the importance of the distinction between the center, bar, and bar-end in the SFE analysis of barred galaxies. Although only massive gas-rich barred galaxies are sufficiently sampled in the current data sets, the increase of CO and SFR data with higher resolution and sensitivity to resolve these environments in the future will enable us to comprehensively understand the relationship between the evolution of the host barred galaxies (i.e., the location in the $M_{\rm star}$ vs. sSFR diagram) and changes in the SFE within galaxies.

\newpage

\begin{acknowledgments}
We would like to thank an anonymous reviewer for useful comments. We thank Kazuyuki Muraoka and Kana Morokuma-Matsui for a fruitful discussion.
F.M., F.E., K.O., Y.F., and A.H. are supported by JSPS KAKENHI grant No. JP21J00108, JP17K14259, JP19K03928, JP22K20387, and JP19K03923, respectively.
F.M. was also supported by the ALMA Japan Research Grant of NAOJ ALMA Project, NAOJ-ALMA-266.
This work is based on COMING and the Nobeyama Atlas of Nearby Spiral Galaxies, which are both legacy programs of the Nobeyama 45 m radio telescope, which is operated by NRO, a branch of National Astronomical Observatory of Japan (NAOJ). 
This work made use of HERACLES, ``The HERA CO-Line Extragalactic Survey'' \citep{leroy_heracles_2009} with the IRAM 30 m telescope. IRAM is supported by INSU/CNRS (France), MPG (Germany), and IGN (Spain).
This paper makes use of the following ALMA data:
ADS/JAO.ALMA 
\#2011.0.00004.SV,
\#2013.1.01161.S, 
\#2015.1.00121.S, 
\#2015.1.00925.S, 
\#2015.1.00956.S, 
\#2016.1.00386.S, 
\#2017.1.00129.S,  
\#2017.1.00392.S, 
\#2017.1.00886.L, 
\#2018.A.00062.S,  
\#2018.1.01651.S, 
\#2019.2.00052.S, 
\#2019.1.00722.S.
Part of these projects have been processed as the PHANGS-ALMA CO (2–-1) survey.
ALMA is a partnership of ESO (representing its member states), NSF (USA), and NINS (Japan), together with NRC (Canada), MOST and ASIAA (Taiwan), and KASI (Republic of Korea), in cooperation with the Republic of Chile. The Joint ALMA Observatory is operated by ESO, AUI/NRAO, and NAOJ.

Data analysis was in part carried out on the Multi-wavelength Data Analysis System operated by the Astronomy Data Center (ADC), NAOJ.
This research also made use of APLpy, an opensource plotting package for Python \citep{Robitaille_2012}.
We acknowledge the usage of the HyperLeda database (\url{http://leda.univ-lyon1.fr}). 
We would like to thank Editage (www.editage.com) for English language editing.

\end{acknowledgments}

\vspace{5mm}
\facilities{ALMA, IRAM 30m, NRO 45m, GALEX, WISE}

\software{CASA \citep{mcmullin_casa_ASPCS}, Astropy \citep{astropy_2018}, APLpy \citep{Robitaille_2012}}, NnumPy \citep{Numpy_harris2020array}, SciPy \citep{Virtanen_scipy}, astroquery \citep{astroquery_2019}

\appendix

\section{Atlas of $\Sigma_{\rm SFR}$, $\Sigma_{\rm mol}$, and environmental mask}  \label{ap: atlas}

Figure~\ref{fig:FoV 1} is the same as Figure~\ref{fig:FoV}, but for all long bar sample galaxies.
We also show the $f_{\rm bar}(\Sigma_{\rm mol} \geqq 5)$ and surface density corresponding to $3\sigma$ upper limits of CO(1--0) and CO(2--1) cubes. Here, the velocity width is assumed to be $20~{\rm km~s^{-1}}$.
Note that we display the region with $\Sigma_{\rm mol} \geqq 5~M_\odot~\rm pc^{-2}$ in the $\Sigma_{\rm mol}$ maps.

\begin{figure*}[p!]
\begin{center}
\includegraphics[width=\hsize]{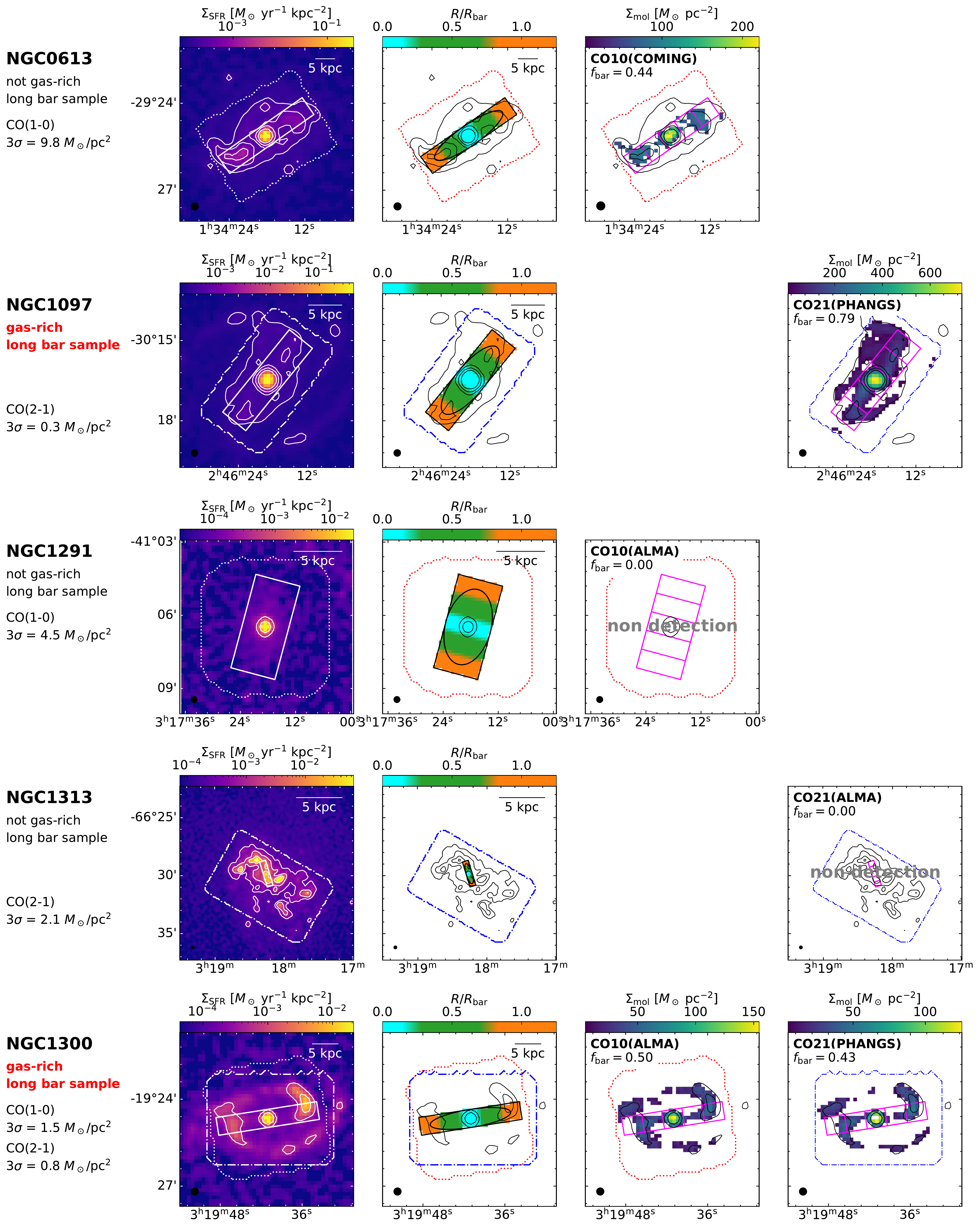}
\caption{Same as Figure~\ref{fig:FoV}, but for all long bar sample galaxies.}
\label{fig:FoV 1}
\end{center}
\end{figure*}

\begin{figure*}[p!]
\addtocounter{figure}{-1} 
\begin{center}
\includegraphics[width=\hsize]{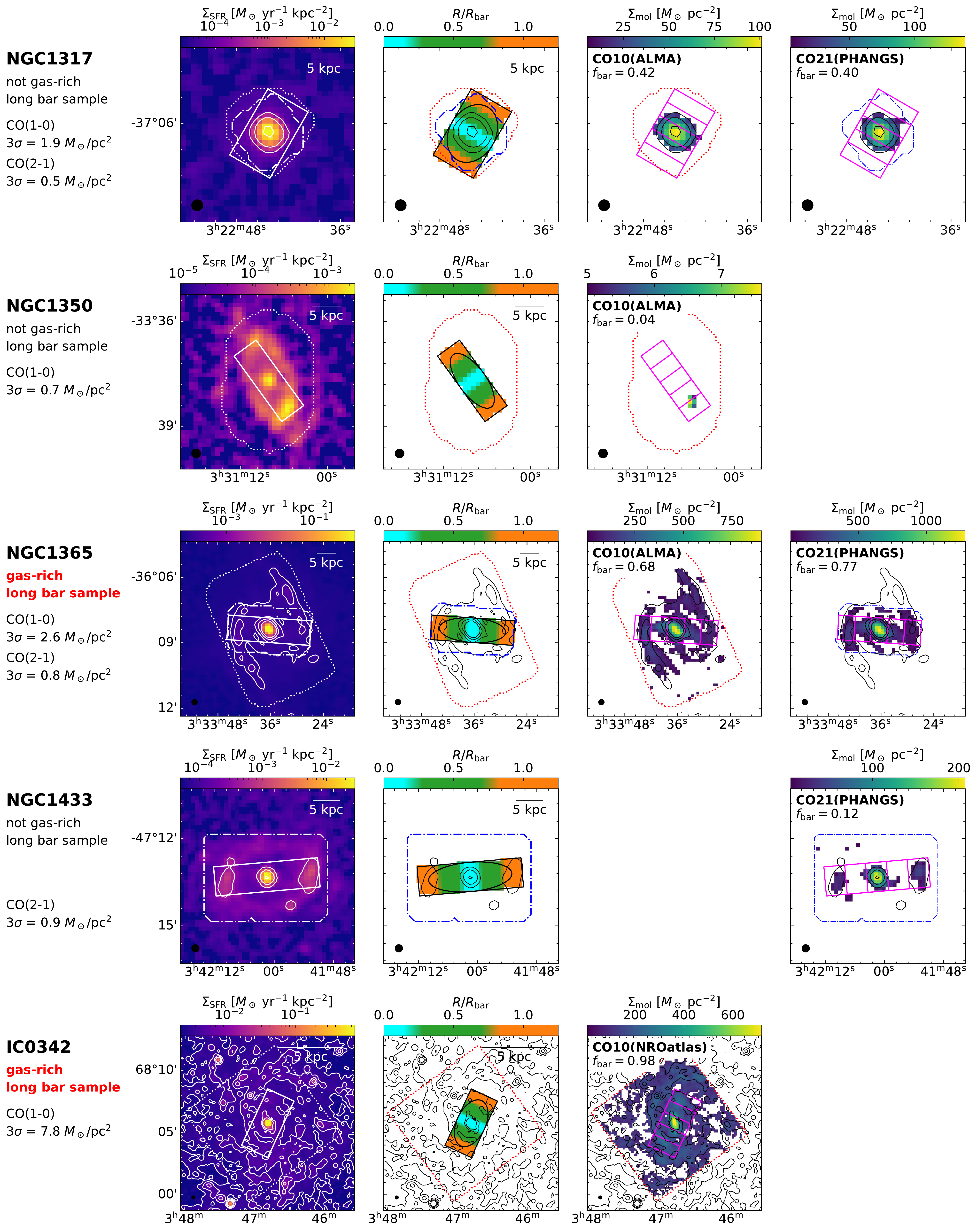}
\caption{(Continued)}
\label{fig:FoV 2}
\end{center}
\end{figure*}

\begin{figure*}[p!]
\addtocounter{figure}{-1} 
\begin{center}
\includegraphics[width=\hsize]{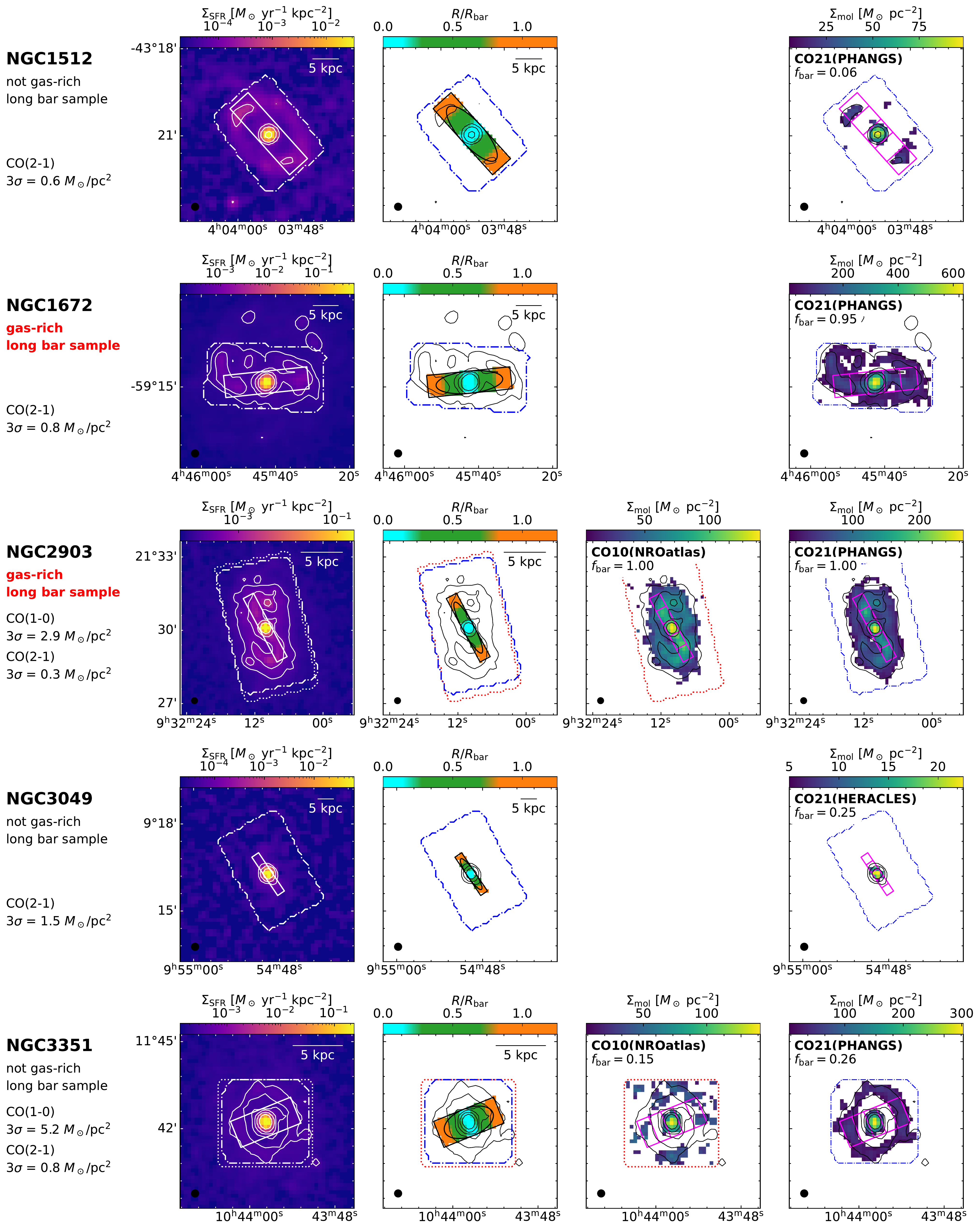}
\caption{(Continued)}
\label{fig:FoV 3}
\end{center}
\end{figure*}

\begin{figure*}[p!]
\addtocounter{figure}{-1} 
\begin{center}
\includegraphics[width=\hsize]{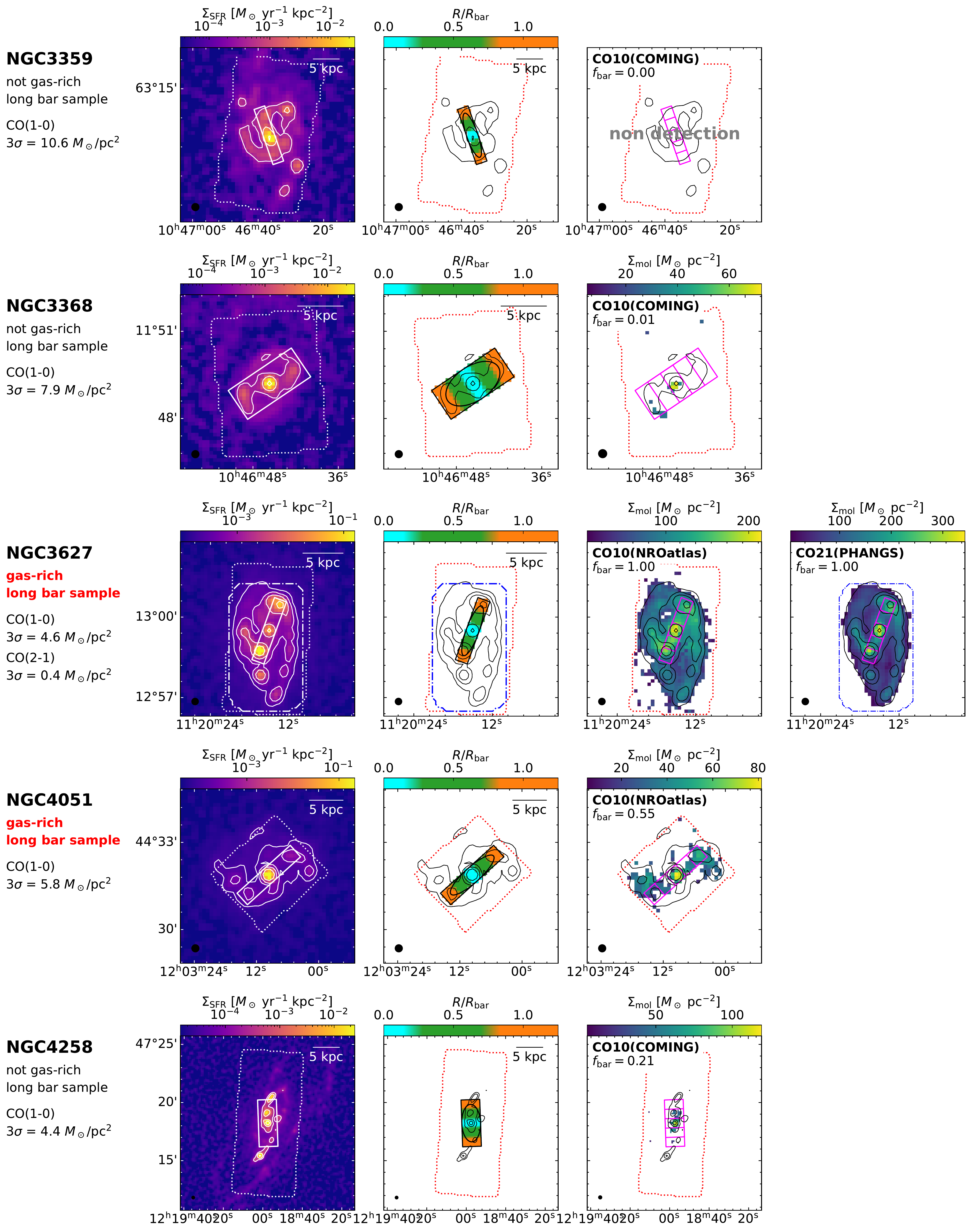}
\caption{(Continued)}
\label{fig:FoV 3}
\end{center}
\end{figure*}

\begin{figure*}[p!]
\addtocounter{figure}{-1} 
\begin{center}
\includegraphics[width=\hsize]{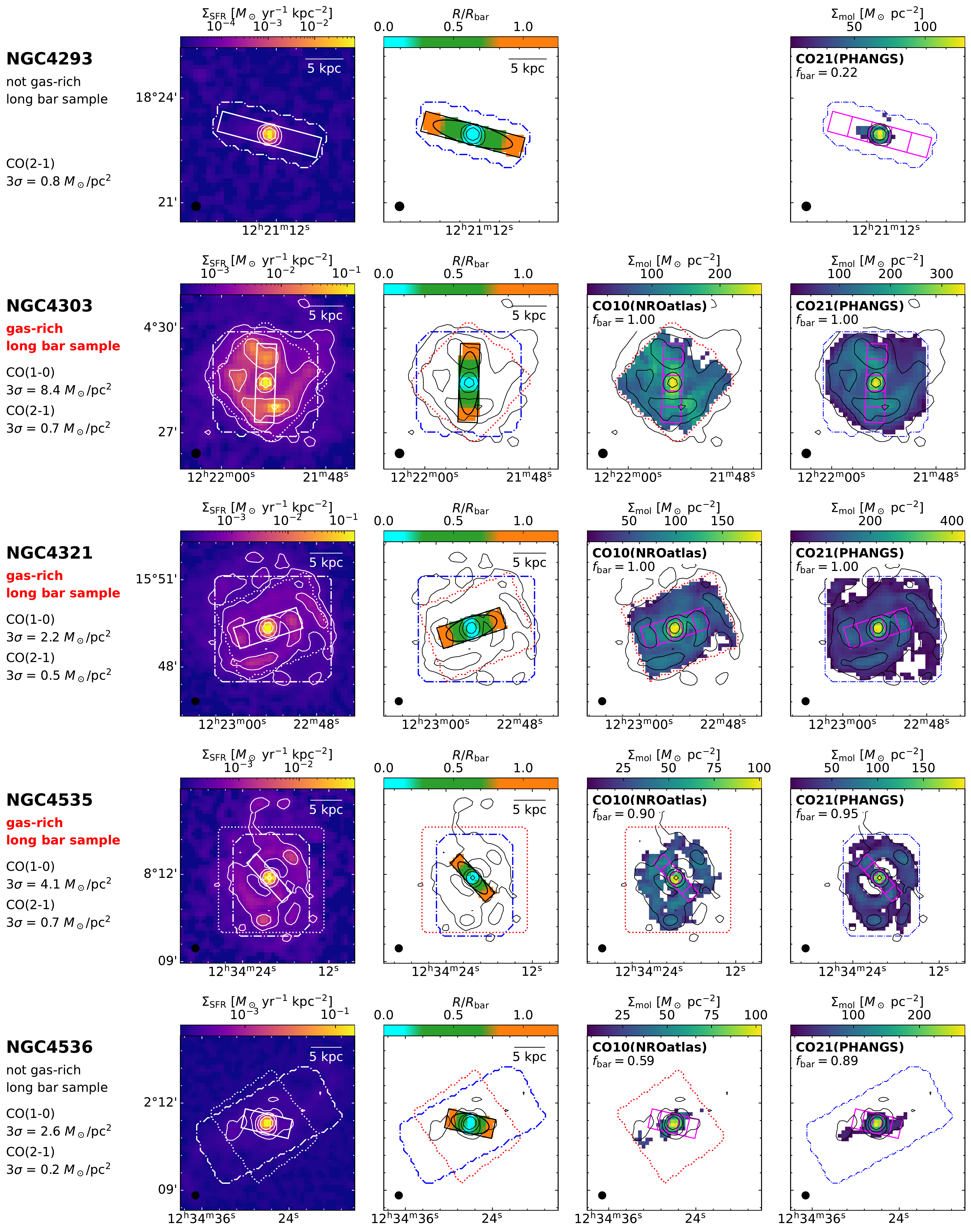}
\caption{(Continued)}
\label{fig:FoV 3}
\end{center}
\end{figure*}

\begin{figure*}[p!]
\addtocounter{figure}{-1} 
\begin{center}
\includegraphics[width=\hsize]{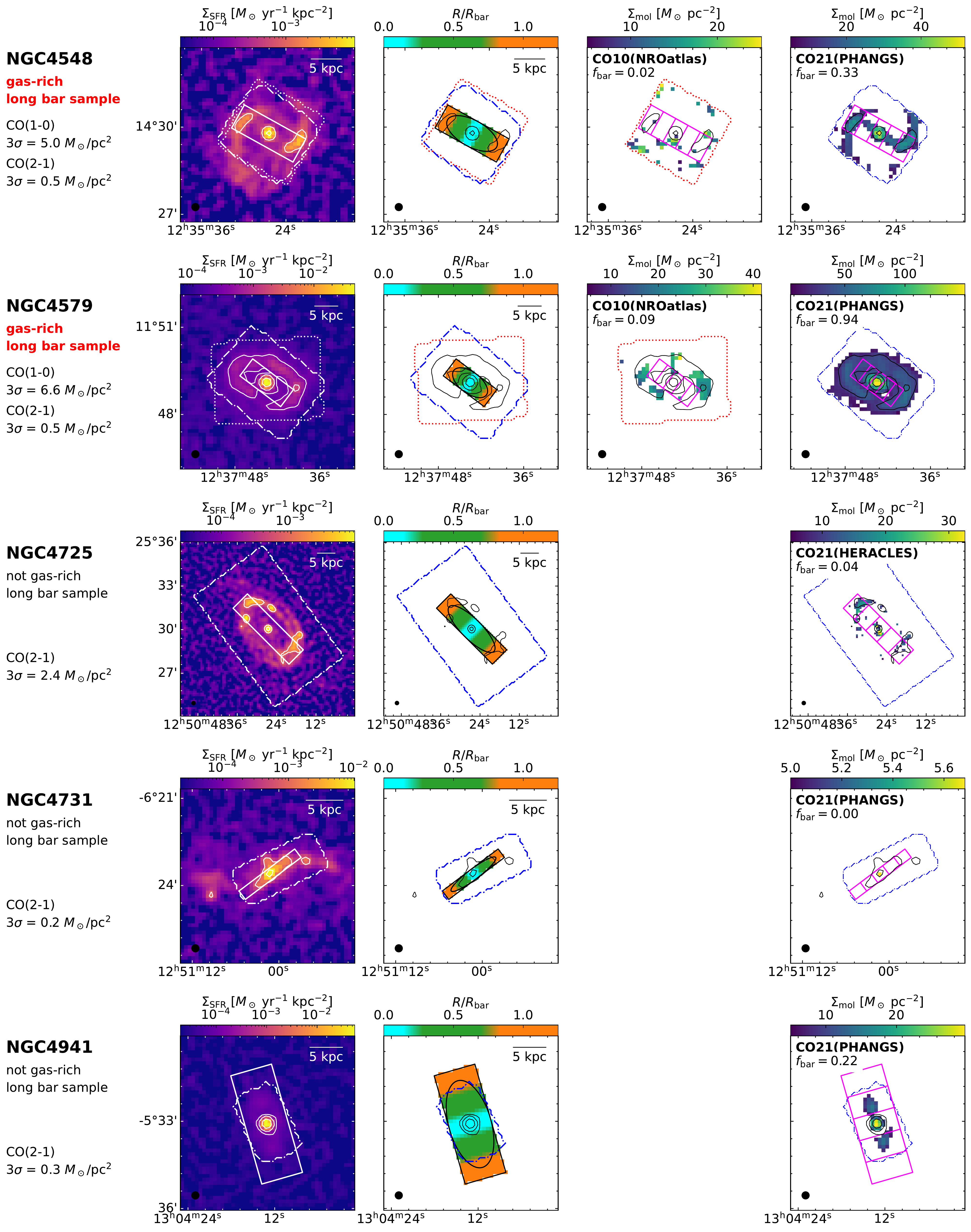}
\caption{(Continued)}
\label{fig:FoV 3}
\end{center}
\end{figure*}

\begin{figure*}[p!]
\addtocounter{figure}{-1} 
\begin{center}
\includegraphics[width=\hsize]{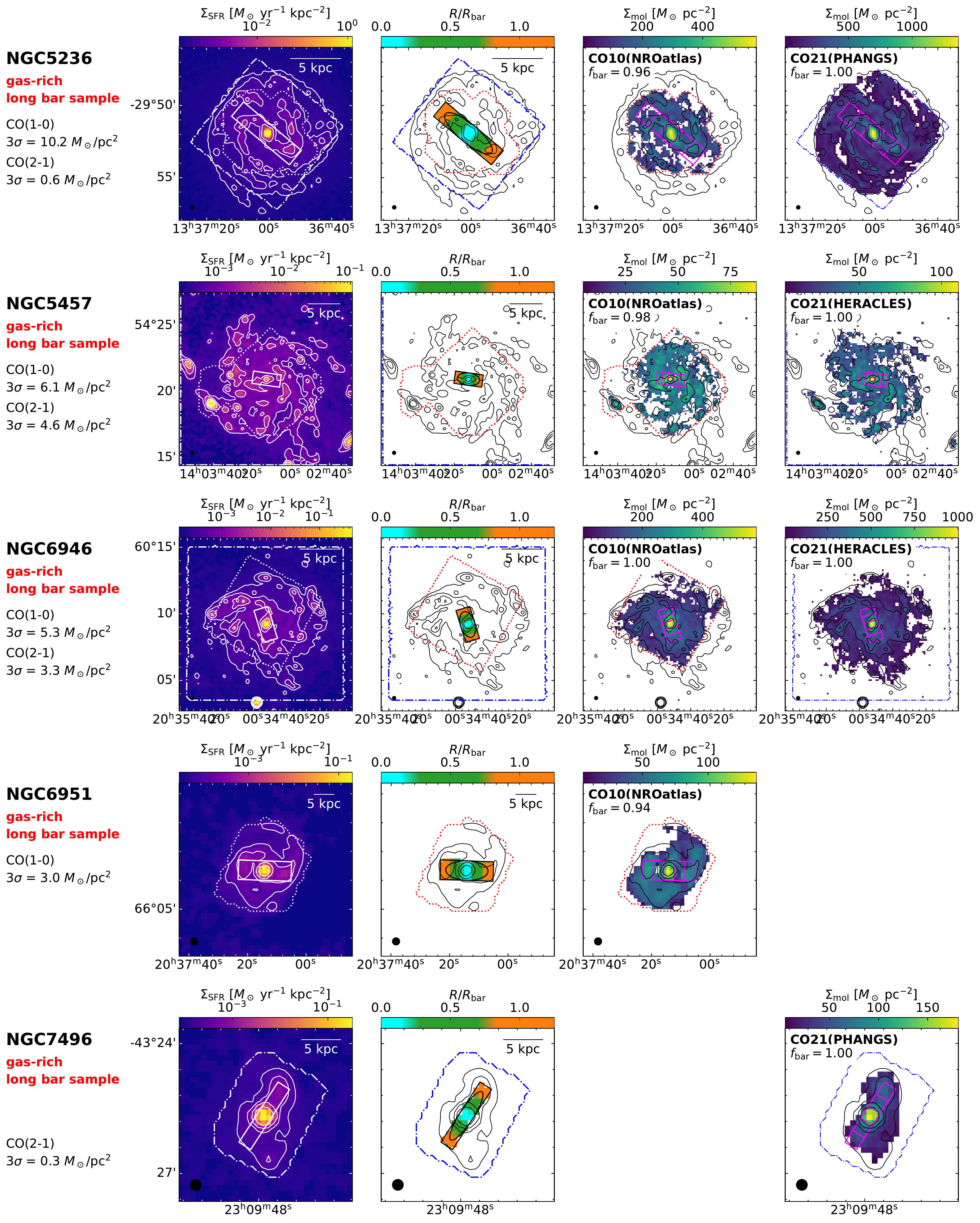}
\caption{(Continued)}
\label{fig:FoV 3}
\end{center}
\end{figure*}

\bibliographystyle{aasjournal}
\bibliography{Reference_SFE}

\end{document}